\documentclass[useAMS,usenatbib,usegraphicx]{mn2e} \def\kms{km s$^{-1}$}

\usepackage[T1]{fontenc}
\usepackage{aecompl}
\title[The Minute Cadence Survey] {The DECam Minute Cadence Survey I.}

\author[C. Belardi et al.]
{Claudia Belardi$^{1,2}$,
Mukremin Kilic$^{1}$,
Jeffrey A. Munn$^{3}$, 
A. Gianninas$^{1}$, 
\newauthor Sara D. Barber$^{1,6}$,
Arjun Dey$^{4}$,  
Peter B. Stetson$^{5}$\\
$^{1}$Homer L. Dodge Department of Physics and Astronomy, Univesity of Oklahoma,
440 W. Brooks St., Norman, OK 73019, USA\\ 
$^{2}$Department of Physics and Astronomy, University of Leicester, University
Road, Leicester LE1 7RH, UK\\
$^{3}$US Naval Observatory, Flagstaff Station, 10391 W. Naval Observatory Road,
Flagstaff, AZ 86005, USA\\ 
$^{4}$National Optical Astronomy Observatory, 950 N. Cherry Ave., Tucson, AZ 85719, USA\\ 
$^{5}$Dominion Astrophysical Observatory, NRC-Herzberg, 5071 West Saanich Road,
Victoria, BC V9E 2E7, Canada\\ 
$^{6}$American Institute of Physics Congressional Fellow\\
}

\begin{document}

\maketitle

\begin{abstract} 

We present the first results from a minute cadence survey of a
three square degree field obtained with the Dark Energy Camera.  We imaged part
of the Canada-France-Hawaii Telescope Legacy Survey area over eight
half-nights. We use the stacked images to identify 111 high proper motion white
dwarf candidates with $g\leq24.5$ mag and search for eclipse-like events and other sources
of variability. We find a new $g=20.64$ mag pulsating ZZ Ceti star with
pulsation periods of 11-13 min. However, we do not find any transiting planetary companions
in the habitable zone of our target white dwarfs. Given the probability
of eclipses of 1\% and our observing window from the ground, the non-detection
of such companions in this first field is not surprising. Minute cadence DECam
observations of additional fields will provide stringent constraints on the
frequency of planets in the white dwarf habitable zone.
 
\end{abstract}


\section{Introduction}

Transient surveys like the Palomar Transient Factory \citep{rau09}, Panoramic
Survey Telescope \& Rapid Response System Medium Deep Fields
\citep{kaiser10,tonry12}, Dark Energy Survey Supernova Fields
\citep{flaugher05,bernstein12}, Sloan Digital Sky Survey Stripe 82
\citep{ivezic07}, Catalina surveys \citep{drake09}, as well as microlensing surveys
like the Massive Compact Halo Objects project \citep{alcock00} and the Optical
Gravitational Lensing Experiment \citep{udalski03} have targeted large areas of
the sky with hour to day cadences to identify variable objects like supernovae,
novae, Active Galactic Nuclei, cataclysmic variables, eclipsing and contact
binaries, and microlensing events. 

Several exoplanet surveys, e.g., the Wide Angle Search for Planets (WASP) and
Hungarian-made Automated Telescope Network (HATNet), have used a number of small
cameras or telescopes to obtain $\sim$few min cadence photometry on a large
number of stars, providing 1\% photometry for stars brighter than 12 mag. Yet
other transient surveys targeted specific types of stars, like M dwarfs for the
MEarth project, to look for exoplanets around them. The largest exoplanet
survey so far, the Kepler mission, provided short cadence ($\approx$1 min) data
for 512 targets in the original mission, and the ongoing K2 mission is adding
several dozen more short cadence targets for each new field observed. 
One of the unusual findings from the Kepler/K2 mission includes
an exciting discovery of
a disintegrating planetesimal around the dusty white dwarf WD 1145+017 in a 4.5 h orbit
\citep{vanderburg15,gansike16,rappaport16}. Such planetesimals around white dwarfs have not been found before
because none of the previous surveys were able to observe a large number of white dwarfs
for an extended period of time. These planetesimals are likely sent closer to the central star
through planet-planet interactions \citep{Jura03, debes12, Vera13}.
Hence, at least some planets must survive the late stages of stellar evolution.

The Large Synoptic Survey Telescope (LSST) will identify about 13 million white dwarfs
and it will provide repeated observations of
the southern sky every 3 days over a period of 10 years. Each LSST visit
consists of two 15 s exposures, reaching a magnitude limit of $g=24.5$ mag.
However, this cadence is not optimum for identifying sources that vary on minute
timescales.  Here we present the first results from a new minute-cadence survey
on the Cerro Tololo 4m Blanco Telescope that reaches the same magnitude limit as
each of the LSST visits. We take advantage of the relatively large field of view of the
Dark Energy Camera \citep[DECam,][]{flaugher05} to perform eight half-night long
observations of individual fields to explore the variability of the sky in
minute timescales.

For this paper, we focus on the 111 high proper motion white
dwarf candidates in our first survey field. We describe the details of our
observations and reductions in Section 2. Sections 3 and 4 provide proper motion
measurements and the sample properties. We present the light curves for the variable
white dwarfs in Section 5, and conclude in Section 6.

\section{DECam Data}

\subsection{Observations}

We used DECam mounted on the Blanco 4m Telescope on UT 2014 Feb 2-9
to obtain $g-$band exposures of a three square degree field (corresponding to a single DECam pointing) centred
at Right Ascension $\alpha =$ 9h 3m 2s and Declination $\delta =$ -4d 35m 0s.
Our observations were performed under the NOAO program 2014A-0073.
This field was previously observed by the Canada-France-Hawaii
Telescope Legacy Survey (CFHTLS\footnote{http://www.cfht.hawaii.edu/Science/CFHLS})
between 2003 and 2008, and is part of the CFHTLS Wide 2 field, which is a 25 square
degree field with MegaCam $ugriz$ photometry available. The earlier MegaCam data provide
the first epoch for our proper motion measurements.

DECam consists of a grid of 62 CCDs, each with size $2048 \times 4096$ pixels
and platescale $0.263 \arcsec$ per pixel.
However, two of the CCDs were not functioning during our observing run, reducing
the number of usable CCDs to 60.

All of our observations were obtained during the second
half-of-the night, resulting in 4 hour long observing windows each night.
The airmass of the target field ranged from 1.1 to 2.5 with a median airmass of 1.3
for the entire run. The $g-$band seeing ranged from 0.97 to $2.08 \arcsec$, with a median
seeing of $1.23 \arcsec$. Given the change in seeing, the individual
exposure times ranged from 70 to 90 s, leading to an overall cadence of
$\approx$90 to 110 s due to the $\approx$20 s read-out time of the camera. 
These exposure times were chosen to obtain S/N$\geq5$ photometry of
targets brighter than $g=24.5$ AB mag under the different seeing conditions.
We obtained a total of 1041 DECam images of this field.

\subsection{Data Reduction}

We downloaded the reduced and calibrated DECam images from the NOAO
Science Archive. Our images were processed through the NOAO
``Community Pipeline'' (version 3.0.2), as described in the NOAO Data
Handbook\footnote{http://ast.noao.edu/sites/default/files/NOAO\_DHB\_v2.2.pdf}
(2015). The pipeline uses calibration exposures taken during the observing run,
such as biases and dome-flats, to remove the instrumental signature, and applies astrometric
and photometric calibrations. The dark current in DECam is extremely low, and no dark
correction is applied in the pipeline.

The 2MASS\footnote{http://www.ipac.caltech.edu/2mass/} source positions are used
as a reference to perform the initial astrometric correction. Initial photometric calibration
and estimates of the zero-point magnitudes for the science
images were obtained by comparing the DECam instrumental brightnesses of field stars
to their published magnitudes in the USNO-B1 catalog \citep{Mone03}.

\section{High Proper Motion White Dwarfs} 

\subsection{Proper Motions}

The CFHT Legacy Survey Wide 2 field is a low-extinction, $E(B-V)=0.02$ mag,
$4.8^{\circ} \times 4.7^{\circ}$ field located at a Galactic latitude of
$l = +26.6 ^{\circ}$. The MegaCam data on this field reach a completeness
limit of $g=25.5$ mag. In order to reach a comparable or better limiting
magnitude, we stack 706 of our DECam images with airmass $<1.5$ and a median
seeing of $1.16 \arcsec$. We stack the images on a chip by chip basis using the IRAF \textit{imcombine} package, and
trim off the bad portions of each image, especially near the chip edges.
 
\begin{figure}
\vspace{-0.2in}
\includegraphics[width=\columnwidth]{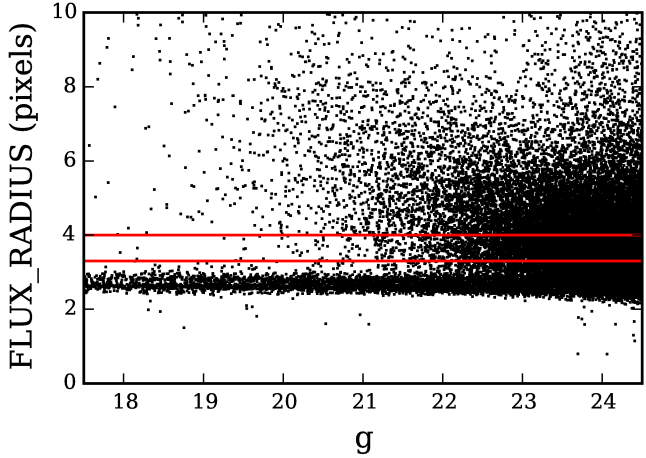}
\caption{Object size (SExtractor's FLUX\_RADIUS) versus $g$ magnitude, measured
on the CFHTLS $g$ image (only 20\% of the objects detected have been
plotted).  The red lines define the band in object size from which
astrometric calibrators were selected, corresponding to barely
resolved galaxies.}
\label{fig:sizemag}
\end{figure}

We used SExtractor \citep{bertin96} to detect objects in the stacked
image, and measured their positions (based on the windowed first order moments of the image profile, XWIN\_IMAGE and YWIN\_IMAGE) and instrumental magnitudes.  These
were matched to the CFHT Legacy Survey (CFHTLS) Wide 2 Field from the
Terapix T0007 data release.  In order to derive absolute proper
motions, we calibrated the DECam astrometry against the CFHTLS using
barely resolved galaxies.  Figure \ref{fig:sizemag} plots object size measured on the
CFHTLS $g$ image (SExtractor's FLUX\_RADIUS) versus magnitude.
Astrometric calibrators were selected to have $3.3 <$ FLUX\_RADIUS $<
4.0$ (indicated by the red lines in Figure \ref{fig:sizemag}), which avoids stars (the
locus of points with FLUX\_RADIUS $\sim$2.7), but limits the calibrators to
barely resolved galaxies, for which accurate centroids can be
measured.  Calibrators were further required to be clean (based on SExtractor flags) sources with $17.5 < g < 24$. Blended sources were removed by rejecting objects with a neighbour within 2$ \arcsec$. For each CCD, separate affine
transformations in right ascension and declination were then fit to
the offsets between the CFHTLS and DECam positions.  Proper motions
were then derived by differencing the CFHTLS and recalibrated DECam
positions.

\begin{figure}
\vspace{-0.2in}
\includegraphics[width=\columnwidth]{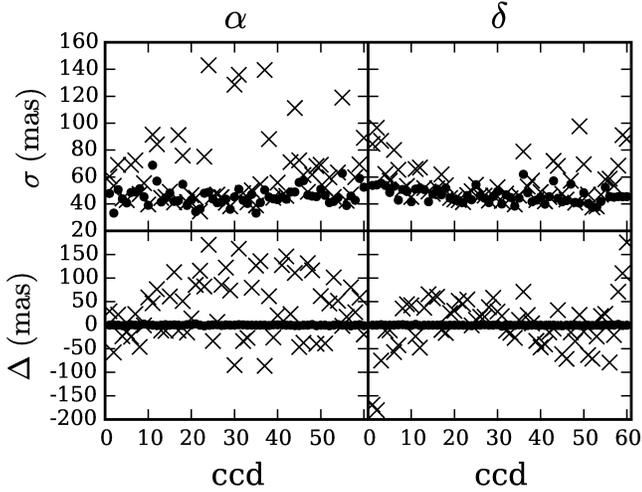}
\caption{The rms (top panels) and mean (bottom panels) differences between the
CFHTLS and DECam positions in right ascension (left panels) and
declination (right panels).  The crosses are the differences before
the recalibration, and the filled circles after the recalibration.}
\label{fig:calibration}
\end{figure}

Figure \ref{fig:calibration} shows the mean and rms
differences between the CHFTLS and DECam positions for each CCD before and
after the calibrations. 
Prior to calibrations, the average offsets for each CCD are of order
20 - 150 mas, with RMS of order 50 - 150 mas. After astrometric calibrations,
the RMS for each CCD is reduced to $\approx$ 50 mas.

To estimate proper motion errors as a function of magnitude,
we use the RMS of proper motion distributions of barely resolved
galaxies at the bright end. We do not use the relatively bright stars
for this estimate since the real stellar motions inflate
the proper motion RMS. At the faint end, the errors in the centroids of the
galaxies inflate the rms, so we use the rms of the proper motion
distributions of the stars, which are mostly unaffected by real
stellar motions.
Proper motion errors are roughly 5 mas yr$^{-1}$ down to $g=24.5$ mag. We find
$\sigma_{\mu} = 4.47$ mas yr$^{-1}$ for objects brighter than
$g=21.5$ mag, and $\sigma_{\mu}=4.69+0.183 \times (g-21.5)^2$ for
$g>21.5$ mag. Calibration errors dominate the centroiding errors, even at the faint end.

\begin{figure} 
\includegraphics[width=\columnwidth]{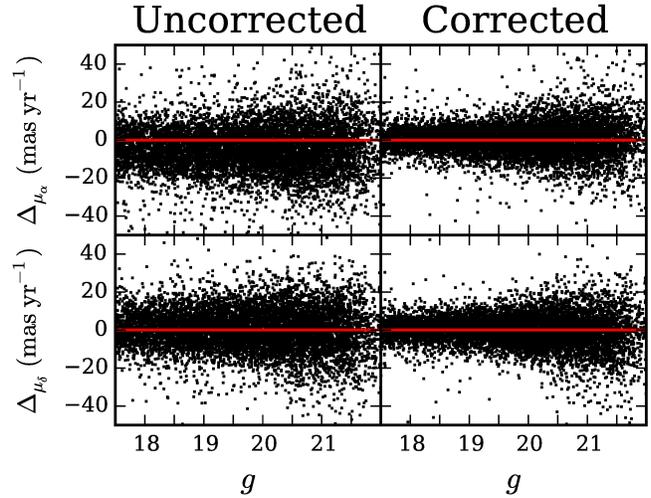} 
\caption{The difference between our proper motions -- in both Right Ascension (top) and
Declination (bottom) -- and the PPMXL catalog. Left panels show the comparison
before the astrometric correction, while the right panels show the same
comparison after the correction. Our proper motions agree with the PPMXL catalog
within 5 mas yr$^{-1}$ for objects brighter than $g=19$ mag.} 
\label{fig:ppmxl}
\end{figure}

We compare our proper motion measurements to that of the PPMXL
catalog ~\citep{Roes10}. Figure \ref{fig:ppmxl} shows this comparison
for both before and after the astrometric calibration of our DECam data.
The corrected proper motions agree well with the PPMXL values, showing an
RMS scatter of 5.3-5.5 mas year$^{-1}$ for stars brighter than $g=19$ mag.

\subsection{The Reduced Proper Motion Diagram}

\begin{figure*}
\includegraphics[width=0.6\textwidth]{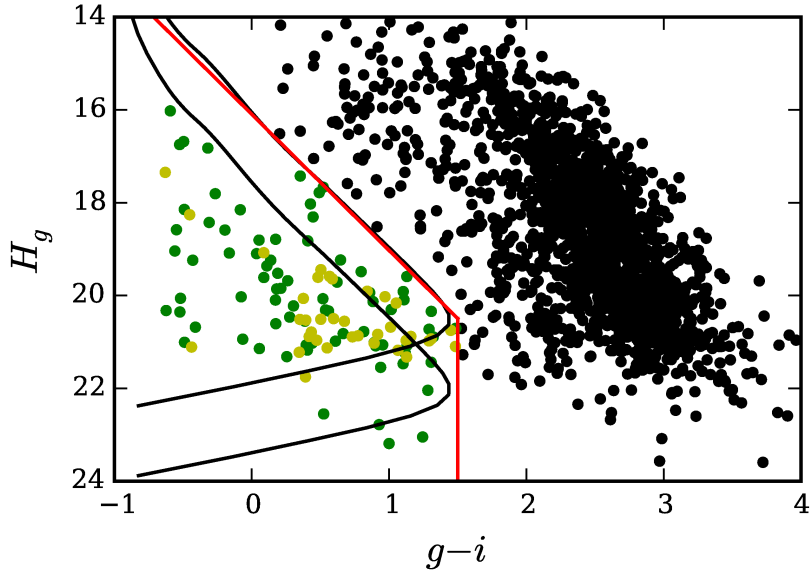}
\caption{The Reduced Proper Motion diagram for stars in the first DECam field. White dwarf
cooling curves for tangential velocities of 20 and 40 \kms\ are shown as solid
lines. The red line delineates the area of interest for white dwarfs; we
visually inspected each object in this region and classified them as ``good'' (dark
green) and ``maybe'' (yellow) white dwarf candidates.}
\label{fig:rpm}
\end{figure*}

Reduced proper motion, defined as $H = m + 5 \log{\mu} + 5 =
M + 5 \log{V_{\rm tan}} - 3.379$, can be used as a proxy for absolute
magnitude for samples with similar kinematics. It also provides a relatively
clean selection of different populations of stars, including white dwarfs and
halo subdwarfs \citep{kilic06,kilic10}.

To generate a clean sample of high proper motion stars, we
first apply the cut $2.1 <$ FLUX\_RADIUS $< 3.3$ on the CFHTLS $g$ image,
corresponding to the stellar locus in Figure 1.  To remove the
considerable contamination from galaxies at the faint end, we require
a clean, significant proper motion, for which we adopt the following
criteria: 1) a one-to-one match between the DECam and CFHT images; 2)
a matching distance of less than $4\arcsec$, to avoid mismatches; 3) no neighbouring
object within $2\arcsec$, to avoid blends; 4) ellipticity measured on the DECam
image less than 0.2; and 5) a total proper motion of $\mu > 20$ mas yr$^{-1}$,
corresponding to roughly a $4\sigma$ detection.

Figure \ref{fig:rpm} shows the reduced proper motion diagram for our first
DECam field, along with white dwarf evolutionary tracks for tangential
velocities of 20 and 40 km s$^{-1}$. The red line marks the boundaries
for our white dwarf selection region. We visually inspected all objects in this region on both the CFHTLS and DECam images, and classified the likelihood that the proper motion is real.  Those objects whose CFHTLS or DECam centroids were clearly wrong (due to such effects as blends with nearby bright stars, false detections due to cosmic rays or diffraction spikes, etc), or whose image profiles are clearly non-stellar, have been rejected and are not plotted.  The remaining objects were classified as either ``good'' or ``maybe''.  ``Good'' objects have clean stellar profiles, and their proper motions are  likely real.  Objects classified as ``maybe'' are not obviously wrong (e.g., a clear blend with a nearby star), but one or both centroids may be incorrect.  Typically one of three concerns was evident on the CFHTLS and/or DECam images: 1) the image profile is slightly asymmetric; 2) there is a nearby neighbor, though not obviously so close as to affect the centroid; or 3) the object is faint enough that it may not be possible to measure an accurate centroid.  We suspect that the vast majority of the ``maybes'' have unreliable proper motions, and thus are not white dwarfs (and subsequent analysis in the paper will support this suspicion), however we carry them through the analysis as we also suspect a few may have reliably measured proper motions, and to include them in our photometric analysis in case they show variability or eclipses.  We identify 78 good white dwarf candidates and 33 ``maybes''.

\begin{table*} 
\centering 
\scriptsize 
\caption{Astrometry and photometry of the good white dwarf candidates in the first field.}
\begin{tabular}{llccccccrr} 
\hline 
Target & RA  &  Dec & {u} & {g} & {r} & {i} & {z} & $\mu_{RA}$ &  $\mu_{Dec}$  \\ Name & (deg) &  (deg) & (mag) & (mag) &
(mag) & (mag) & (mag) & \multicolumn{2}{c}{(mas/year)} \\ 
\hline 
WD1  & 135.428180 & -3.683920 & 24.74$\pm$0.13 & 24.12$\pm$0.04 & 23.85$\pm$0.06 & 23.46$\pm$0.06 & 23.60$\pm$0.13 & -19.8 & 9.2 \\
WD2  & 135.977576 & -3.644388 & 22.42$\pm$0.02 & 22.35$\pm$0.01 & 22.55$\pm$0.02 & 22.91$\pm$0.05 & 23.15$\pm$0.13 & 8.5 & -20.1 \\
WD5  & 135.564771 & -3.835989 & 18.62$\pm$0.00 & 18.45$\pm$0.00 & 18.75$\pm$0.00 & 19.05$\pm$0.00 & 19.32$\pm$0.00 & -25.1 & -20.9 \\
WD6  & 135.783438 & -3.789615 & 21.17$\pm$0.01 & 20.79$\pm$0.01 & 20.97$\pm$0.01 & 21.10$\pm$0.01 & 21.27$\pm$0.03 & -12.1 & -31.4 \\
WD7  & 136.305058 & -3.806940 & 25.50$\pm$0.30 & 24.09$\pm$0.05 & 22.96$\pm$0.04 & 22.77$\pm$0.05 & 22.39$\pm$0.04 & 16.2 & 16.2 \\
WD9  & 135.243361 & -4.059065 & 22.70$\pm$0.03 & 22.36$\pm$0.01 & 22.30$\pm$0.02 & 22.33$\pm$0.02 & 22.55$\pm$0.07 & 10.5 & -19.6 \\
WD10 & 135.281163 & -4.005761 & 24.09$\pm$0.10 & 23.53$\pm$0.04 & 23.24$\pm$0.05 & 23.09$\pm$0.05 & 23.43$\pm$0.15 & -27.9 & 14.9 \\
WD12 & 135.864665 & -4.025207 & 21.42$\pm$0.01 & 21.09$\pm$0.00 & 21.29$\pm$0.01 & 21.58$\pm$0.01 & 21.91$\pm$0.04 & -18.3 & -18.1 \\
WD14 & 136.240070 & -3.939055 & 24.12$\pm$0.10 & 23.80$\pm$0.06 & 23.77$\pm$0.08 & 24.21$\pm$0.21 & 23.52$\pm$0.15 & -22.5 & -7.8 \\
WD15 & 135.126251 & -4.112492 & 24.98$\pm$0.20 & 24.45$\pm$0.11 & 24.00$\pm$0.12 & 23.84$\pm$0.10 & 23.40$\pm$0.16 & 16.8 & 11.3 \\
WD16 & 135.127629 & -4.107993 & 26.27$\pm$0.55 & 24.18$\pm$0.07 & 23.18$\pm$0.05 & 22.69$\pm$0.03 & 22.58$\pm$0.07 & 15.9 & 12.5 \\
WD17 & 135.497429 & -4.186616 & 23.81$\pm$0.06 & 22.73$\pm$0.02 & 22.14$\pm$0.02 & 21.94$\pm$0.01 & 21.81$\pm$0.03 & -21.1 & -7.5 \\
WD18 & 135.542934 & -4.120156 & 24.62$\pm$0.13 & 23.30$\pm$0.03 & 22.70$\pm$0.03 & 22.44$\pm$0.02 & 22.36$\pm$0.05 & 11.8 & -17.8 \\
WD19 & 135.494243 & -4.103245 & 26.00$\pm$0.44 & 24.39$\pm$0.08 & 23.75$\pm$0.07 & 23.39$\pm$0.05 & 23.34$\pm$0.12 & -55.7 & -14.1 \\
WD20 & 136.128472 & -4.189202 & 20.28$\pm$0.00 & 19.94$\pm$0.00 & 20.23$\pm$0.00 & 20.46$\pm$0.01 & 20.75$\pm$0.01 & -4.9 & -22.5 \\
WD21 & 136.513464 & -4.189520 & 22.39$\pm$0.02 & 21.97$\pm$0.01 & 22.26$\pm$0.02 & 22.40$\pm$0.04 & 22.73$\pm$0.06 & -27.7 & -6.2 \\
WD22 & 136.518411 & -4.175147 & 21.57$\pm$0.01 & 21.26$\pm$0.01 & 21.32$\pm$0.01 & 21.34$\pm$0.01 & 21.48$\pm$0.02 & -6.1 & -23.2 \\
WD23 & 134.839947 & -4.279327 & 24.32$\pm$0.19 & 24.03$\pm$0.06 & 23.98$\pm$0.14 & 23.64$\pm$0.11 & 22.69$\pm$0.11 & 13.9 & 14.7 \\
WD24 & 134.976040 & -4.403990 & 23.52$\pm$0.04 & 23.12$\pm$0.02 & 22.85$\pm$0.03 & 22.72$\pm$0.03 & 22.81$\pm$0.07 & -39.6 & -10.0 \\
WD25 & 135.282237 & -4.397595 & 23.15$\pm$0.04 & 23.30$\pm$0.03 & 23.57$\pm$0.06 & 23.82$\pm$0.08 & 24.49$\pm$0.33 & -5.1 & -21.9 \\
WD28 & 135.564101 & -4.362957 & 22.62$\pm$0.03 & 22.22$\pm$0.01 & 22.05$\pm$0.02 & 21.95$\pm$0.02 & 22.02$\pm$0.05 & -31.1 & 0.6 \\
WD29 & 135.636367 & -4.292325 & 24.48$\pm$0.11 & 23.26$\pm$0.03 & 22.54$\pm$0.02 & 22.24$\pm$0.02 & 22.25$\pm$0.04 & -11.8 & -22.7 \\
WD30 & 136.569797 & -4.347088 & 24.75$\pm$0.19 & 23.79$\pm$0.05 & 23.07$\pm$0.04 & 22.50$\pm$0.04 & 22.14$\pm$0.04 & -17.4 & 17.1 \\
WD31 & 136.515653 & -4.271666 & 22.56$\pm$0.03 & 22.18$\pm$0.01 & 22.15$\pm$0.02 & 22.13$\pm$0.03 & 22.28$\pm$0.04 & -19.4 & -8.4 \\
WD32 & 135.141287 & -4.524212 & 20.26$\pm$0.00 & 19.95$\pm$0.00 & 20.20$\pm$0.00 & 20.44$\pm$0.01 & 20.72$\pm$0.02 & -10.2 & -19.8 \\
WD33 & 135.323189 & -4.562952 & 20.35$\pm$0.00 & 20.04$\pm$0.00 & 20.18$\pm$0.00 & 20.36$\pm$0.00 & 20.59$\pm$0.01 & -21.5 & -7.2 \\
WD34 & 135.471608 & -4.536417 & 22.81$\pm$0.03 & 22.10$\pm$0.01 & 21.64$\pm$0.02 & 21.56$\pm$0.01 & 21.61$\pm$0.03 & -20.3 & -38.9 \\
WD36 & 135.904348 & -4.500070 & 22.87$\pm$0.04 & 21.46$\pm$0.01 & 20.67$\pm$0.01 & 20.33$\pm$0.01 & 20.30$\pm$0.01 & 35.3 & 23.2 \\
WD37 & 136.194402 & -4.439730 & 21.25$\pm$0.01 & 20.86$\pm$0.00 & 20.92$\pm$0.01 & 20.34$\pm$0.01 & 19.93$\pm$0.01 & -23.0 & 0.0 \\
WD39 & 136.260612 & -4.475453 & 21.64$\pm$0.01 & 21.34$\pm$0.01 & 21.20$\pm$0.01 & 21.20$\pm$0.01 & 21.31$\pm$0.02 & -29.6 & -23.7 \\
WD40 & 134.757590 & -4.728047 & 21.80$\pm$0.01 & 21.25$\pm$0.01 & 20.98$\pm$0.01 & 20.83$\pm$0.01 & 20.88$\pm$0.02 & -17.6 & 14.2 \\
WD41 & 134.854198 & -4.631957 & 24.69$\pm$0.17 & 24.38$\pm$0.07 & 24.29$\pm$0.12 & 23.49$\pm$0.08 & 22.85$\pm$0.09 & 17.3 & 12.3 \\
WD42 & 135.214602 & -4.713657 & 21.01$\pm$0.01 & 20.64$\pm$0.00 & 20.64$\pm$0.01 & 20.15$\pm$0.00 & 19.76$\pm$0.01 & 26.3 & -5.3 \\
WD43 & 135.057145 & -4.694766 & 23.31$\pm$0.04 & 22.96$\pm$0.02 & 22.81$\pm$0.03 & 22.78$\pm$0.03 & 22.82$\pm$0.07 & -21.4 & -10.9 \\
WD44 & 135.569948 & -4.658896 & 24.85$\pm$0.16 & 23.58$\pm$0.05 & 22.77$\pm$0.04 & 22.48$\pm$0.03 & 22.53$\pm$0.07 & -31.8 & 20.6 \\
WD45 & 135.569624 & -4.599770 & 24.71$\pm$0.14 & 23.55$\pm$0.04 & 22.74$\pm$0.03 & 22.43$\pm$0.03 & 22.47$\pm$0.06 & -33.3 & -9.4 \\
WD46 & 135.496518 & -4.598295 & 21.59$\pm$0.01 & 21.30$\pm$0.01 & 21.16$\pm$0.01 & 21.19$\pm$0.01 & 21.33$\pm$0.02 & -17.0 & -37.4 \\
WD47 & 135.695222 & -4.654447 & 23.64$\pm$0.05 & 23.27$\pm$0.03 & 23.12$\pm$0.04 & 23.10$\pm$0.04 & 23.32$\pm$0.11 & 18.1 & -14.4 \\
WD48 & 135.750789 & -4.638145 & 22.87$\pm$0.03 & 22.12$\pm$0.01 & 21.68$\pm$0.01 & 21.50$\pm$0.01 & 21.45$\pm$0.02 & -19.0 & -26.9 \\
WD49 & 136.100249 & -4.702578 & 23.60$\pm$0.05 & 22.62$\pm$0.02 & 22.25$\pm$0.03 & 21.51$\pm$0.02 & 21.07$\pm$0.02 & 13.6 & -28.2 \\
WD50 & 136.520183 & -4.701916 & 23.61$\pm$0.05 & 23.01$\pm$0.03 & 22.89$\pm$0.04 & 22.82$\pm$0.05 & 23.10$\pm$0.15 & -7.2 & -18.7 \\
WD51 & 136.250506 & -4.697482 & 24.54$\pm$0.15 & 24.23$\pm$0.08 & 24.14$\pm$0.11 & 23.41$\pm$0.10 & 23.87$\pm$0.19 & -1.5 & -20.4 \\
WD52 & 136.358869 & -4.670933 & 22.77$\pm$0.03 & 22.47$\pm$0.01 & 22.48$\pm$0.02 & 22.38$\pm$0.03 & 22.53$\pm$0.04 & -26.6 & -2.8 \\
WD53 & 136.251762 & -4.649951 & 23.63$\pm$0.06 & 23.17$\pm$0.02 & 22.96$\pm$0.03 & 22.96$\pm$0.05 & 23.09$\pm$0.08 & -0.2 & -21.6 \\
WD54 & 136.521171 & -4.601962 & 23.73$\pm$0.07 & 23.44$\pm$0.04 & 23.13$\pm$0.04 & 22.55$\pm$0.04 & 22.52$\pm$0.05 & -8.1 & -20.2 \\
WD55 & 136.345247 & -4.718459 & 25.13$\pm$0.22 & 23.17$\pm$0.04 & 22.66$\pm$0.04 & 22.07$\pm$0.03 & 22.00$\pm$0.06 & 10.5 & 19.7 \\
WD56 & 135.100303 & -4.839009 & 25.81$\pm$0.28 & 24.32$\pm$0.06 & 23.65$\pm$0.06 & 23.42$\pm$0.06 & 23.33$\pm$0.11 & -22.0 & 5.2 \\
WD59 & 135.214752 & -4.839602 & 21.99$\pm$0.01 & 21.26$\pm$0.01 & 21.17$\pm$0.01 & 20.99$\pm$0.01 & 20.69$\pm$0.01 & -62.5 & -29.8 \\
WD61 & 135.513533 & -4.894903 & 24.25$\pm$0.07 & 23.18$\pm$0.02 & 22.23$\pm$0.02 & 21.88$\pm$0.02 & 21.78$\pm$0.03 & -18.5 & -19.7 \\
WD62 & 135.613412 & -4.776378 & 20.73$\pm$0.01 & 20.35$\pm$0.00 & 20.08$\pm$0.00 & 19.99$\pm$0.00 & 20.01$\pm$0.01 & -24.0 & -10.2 \\
WD63 & 135.245430 & -5.040763 & 25.20$\pm$0.22 & 23.82$\pm$0.05 & 23.93$\pm$0.12 & 24.44$\pm$0.19 & 23.68$\pm$0.23 & 14.4 & -13.9 \\
WD64 & 135.232479 & -5.034223 & 23.61$\pm$0.04 & 23.20$\pm$0.03 & 22.86$\pm$0.03 & 22.80$\pm$0.04 & 22.89$\pm$0.09 & -30.7 & 21.6 \\
WD65 & 135.786908 & -5.059206 & 24.45$\pm$0.10 & 24.33$\pm$0.08 & 23.76$\pm$0.08 & 23.19$\pm$0.05 & 22.84$\pm$0.09 & -7.0 & -20.6 \\
WD66 & 135.788858 & -4.970007 & 24.32$\pm$0.09 & 23.60$\pm$0.04 & 23.21$\pm$0.05 & 23.08$\pm$0.05 & 23.17$\pm$0.12 & -54.3 & -29.3 \\
WD67 & 135.800417 & -4.946950 & 23.55$\pm$0.04 & 23.23$\pm$0.03 & 23.05$\pm$0.04 & 23.06$\pm$0.05 & 23.20$\pm$0.11 & -12.1 & -27.3 \\
WD68 & 135.806544 & -4.936307 & 20.64$\pm$0.01 & 20.34$\pm$0.00 & 20.45$\pm$0.01 & 20.60$\pm$0.01 & 20.78$\pm$0.02 & -22.9 & -21.2 \\
WD69 & 136.022024 & -5.061150 & 20.63$\pm$0.01 & 20.32$\pm$0.00 & 20.58$\pm$0.00 & 20.85$\pm$0.01 & 21.08$\pm$0.02 & 11.4 & -100.8 \\
WD73 & 135.452818 & -5.216489 & 18.39$\pm$0.00 & 18.22$\pm$0.00 & 18.52$\pm$0.00 & 18.77$\pm$0.00 & 19.05$\pm$0.00 & 35.8 & -112.5 \\
WD74 & 135.536843 & -5.096396 & 25.59$\pm$0.28 & 24.10$\pm$0.06 & 23.43$\pm$0.06 & 23.15$\pm$0.05 & 23.48$\pm$0.16 & -22.5 & -10.2 \\
WD75 & 135.839425 & -5.203433 & 23.73$\pm$0.05 & 23.05$\pm$0.02 & 22.70$\pm$0.02 & 22.53$\pm$0.03 & 22.42$\pm$0.05 & -25.2 & 3.4 \\
WD76 & 135.801031 & -5.154201 & 23.45$\pm$0.04 & 22.87$\pm$0.02 & 22.43$\pm$0.02 & 22.31$\pm$0.02 & 22.23$\pm$0.04 & -30.9 & 3.6 \\
WD77 & 135.906606 & -5.143385 & 24.34$\pm$0.08 & 24.22$\pm$0.07 & 24.81$\pm$0.22 & 24.17$\pm$0.15 & 23.19$\pm$0.14 & 19.4 & -14.4 \\
WD78 & 135.864677 & -5.142195 & 21.99$\pm$0.02 & 21.41$\pm$0.01 & 21.03$\pm$0.01 & 20.97$\pm$0.01 & 21.07$\pm$0.02 & 23.9 & 1.5 \\
WD79 & 136.165377 & -5.132097 & 22.70$\pm$0.02 & 22.58$\pm$0.02 & 22.37$\pm$0.03 & 21.93$\pm$0.02 & 21.47$\pm$0.04 & 21.4 & 1.5 \\
WD80 & 136.059217 & -5.128788 & 24.73$\pm$0.12 & 24.17$\pm$0.06 & 23.91$\pm$0.09 & 23.91$\pm$0.11 & 23.55$\pm$0.17 & -17.9 & 20.1 \\
WD81 & 136.207244 & -5.091935 & 25.65$\pm$0.26 & 24.50$\pm$0.08 & 23.57$\pm$0.06 & 23.25$\pm$0.06 & 23.65$\pm$0.19 & 30.8 & -40.9 \\
WD83 & 135.371489 & -5.369883 & 23.79$\pm$0.05 & 23.42$\pm$0.03 & 23.20$\pm$0.04 & 23.12$\pm$0.05 & 23.25$\pm$0.11 & 19.7 & -11.6 \\
WD84 & 135.436346 & -5.332230 & 21.11$\pm$0.01 & 20.85$\pm$0.00 & 20.91$\pm$0.01 & 21.04$\pm$0.01 & 21.27$\pm$0.02 & -29.8 & 19.1 \\
WD86 & 135.553019 & -5.393839 & 25.49$\pm$0.22 & 24.46$\pm$0.08 & 23.94$\pm$0.08 & 23.53$\pm$0.07 & 23.76$\pm$0.20 & -45.7 & 6.4 \\
WD87 & 135.550518 & -5.278519 & 24.56$\pm$0.09 & 23.67$\pm$0.04 & 23.81$\pm$0.07 & 24.93$\pm$0.24 & 24.80$\pm$0.46 & 49.0 & -81.7 \\
WD88 & 135.893836 & -5.262233 & 25.98$\pm$0.34 & 24.36$\pm$0.07 & 23.51$\pm$0.06 & 23.07$\pm$0.05 & 23.48$\pm$0.17 & 32.5 & -11.5 \\
WD89 & 135.575406 & -5.550305 & 25.61$\pm$0.24 & 24.34$\pm$0.07 & 23.60$\pm$0.06 & 23.21$\pm$0.05 & 23.12$\pm$0.10 & -19.2 & -11.1 \\
WD90 & 135.575378 & -5.446560 & 22.07$\pm$0.02 & 21.74$\pm$0.01 & 21.57$\pm$0.01 & 21.57$\pm$0.02 & 21.66$\pm$0.03 & 3.2 & -25.4 \\
WD94 & 136.124514 & -5.519007 & 22.39$\pm$0.01 & 21.84$\pm$0.01 & 21.53$\pm$0.01 & 21.43$\pm$0.01 & 21.40$\pm$0.03 & 0.1 & -24.9 \\
WD95 & 136.032793 & -5.513574 & 24.15$\pm$0.07 & 23.25$\pm$0.03 & 22.55$\pm$0.02 & 22.31$\pm$0.03 & 22.08$\pm$0.05 & -41.7 & -18.4 \\
WD96 & 135.947007 & -5.493342 & 22.01$\pm$0.01 & 21.69$\pm$0.01 & 21.68$\pm$0.01 & 21.77$\pm$0.02 & 21.78$\pm$0.03 & -43.1 & 17.3 \\
WD97 & 136.152688 & -5.474827 & 21.69$\pm$0.01 & 21.40$\pm$0.01 & 21.43$\pm$0.01 & 21.57$\pm$0.01 & 21.55$\pm$0.03 & 9.6 & -33.0 \\
WD98 & 136.216793 & -5.416903 & 24.53$\pm$0.12 & 24.15$\pm$0.06 & 24.12$\pm$0.10 & 24.21$\pm$0.17 & 24.23$\pm$0.43 & 22.7 & 2.3 \\
\hline 
\end{tabular} 
\end{table*}

\begin{table*} 
\centering 
\scriptsize 
\caption{Astrometry and photometry of the ``maybe'' white dwarf candidates in the first field.}
\begin{tabular}{llccccccrr} 
\hline 
Target & RA  &  Dec & {u} & {g} & {r} & {i} & {z} & $\mu_{RA}$ &  $\mu_{Dec}$ \\
Name & (deg) &  (deg) & (mag) & (mag) & (mag) & (mag) & (mag) & \multicolumn{2}{c}{(mas/year)} \\
\hline 
WD3   & 135.942340 & -3.675134 & 23.59$\pm$0.04 & 22.76$\pm$0.01 & 22.26$\pm$0.02 & 22.19$\pm$0.03 & 22.20$\pm$0.05 & -20.2 & -11.6 \\
WD4   & 136.106871 & -3.639555 & 21.57$\pm$0.01 & 21.57$\pm$0.01 & 21.83$\pm$0.01 & 22.02$\pm$0.02 & 22.38$\pm$0.07 & 7.0 & -20.6 \\
WD8   & 136.300318 & -3.798416 & 24.34$\pm$0.13 & 23.91$\pm$0.06 & 23.26$\pm$0.05 & 23.23$\pm$0.10 & 22.25$\pm$0.09 & 21.2 & 2.6 \\
WD11  & 135.132693 & -3.937421 & 23.97$\pm$0.08 & 23.84$\pm$0.06 & 23.46$\pm$0.07 & 23.24$\pm$0.06 & 23.10$\pm$0.13 & -2.5 & -21.3 \\
WD13  & 135.676317 & -3.974249 & 24.56$\pm$0.24 & 24.30$\pm$0.18 & 22.66$\pm$0.06 & 23.24$\pm$0.12 & 22.28$\pm$0.17 & 5.1 & -23.2 \\
WD26  & 135.258431 & -4.386883 & 24.46$\pm$0.18 & 24.37$\pm$0.12 & 24.27$\pm$0.17 & 23.82$\pm$0.12 & 23.63$\pm$0.25 & -15.4 & -16.4 \\
WD27  & 135.160187 & -4.305303 & 25.08$\pm$0.26 & 24.18$\pm$0.09 & 24.00$\pm$0.12 & 23.41$\pm$0.07 & 23.21$\pm$0.14 & 18.2 & 11.9 \\
WD35  & 135.310004 & -4.532012 & 24.45$\pm$0.11 & 24.41$\pm$0.08 & 24.39$\pm$0.18 & 24.85$\pm$0.21 & 26.20$\pm$1.77 & 7.4 & -20.6 \\
WD38  & 136.217159 & -4.511356 & 23.74$\pm$0.12 & 23.52$\pm$0.06 & 22.99$\pm$0.07 & 23.24$\pm$0.15 & 22.94$\pm$0.14 & 4.5 & 27.4 \\
WD57  & 135.082199 & -4.867261 & 24.99$\pm$0.19 & 24.23$\pm$0.08 & 23.33$\pm$0.06 & 22.78$\pm$0.05 & 22.32$\pm$0.06 & -2.7 & -20.0 \\
WD58  & 135.126308 & -4.819013 & 24.12$\pm$0.17 & 24.30$\pm$0.18 & 19.47$\pm$0.00 & 23.91$\pm$0.25 & 22.98$\pm$0.23 & 30.1 & 7.6 \\
WD60  & 135.369363 & -4.831089 & 26.72$\pm$0.82 & 24.32$\pm$0.08 & 23.63$\pm$0.08 & 23.58$\pm$0.09 & 23.49$\pm$0.17 & -2.6 & 20.5 \\
WD70  & 135.772781 & -5.053292 & 24.79$\pm$0.11 & 24.29$\pm$0.06 & 23.53$\pm$0.05 & 23.17$\pm$0.05 & 23.07$\pm$0.09 & -12.1 & -17.9 \\
WD71  & 135.862551 & -4.998320 & 19.89$\pm$0.00 & 19.87$\pm$0.00 & 20.18$\pm$0.00 & 20.50$\pm$0.01 & 20.72$\pm$0.01 & -21.4 & -22.8 \\
WD72  & 136.049139 & -5.006310 & 23.51$\pm$0.04 & 22.89$\pm$0.02 & 22.54$\pm$0.03 & 22.41$\pm$0.03 & 22.68$\pm$0.08 & -20.4 & 8.2 \\
WD82  & 136.169766 & -5.227046 & 24.17$\pm$0.07 & 24.08$\pm$0.06 & 23.84$\pm$0.08 & 23.65$\pm$0.09 & 23.69$\pm$0.22 & -10.6 & -19.2 \\
WD85  & 135.339775 & -5.283070 & 24.41$\pm$0.11 & 24.21$\pm$0.07 & 24.25$\pm$0.12 & 23.79$\pm$0.10 & 23.23$\pm$0.14 & 18.9 & -10.5 \\
WD91  & 135.594910 & -5.552363 & 24.47$\pm$0.08 & 23.27$\pm$0.03 & 22.51$\pm$0.02 & 22.30$\pm$0.02 & 22.23$\pm$0.04 & -19.6 & -10.9 \\
WD92  & 135.594256 & -5.546409 & 22.51$\pm$0.02 & 22.43$\pm$0.01 & 22.25$\pm$0.02 & 21.92$\pm$0.02 & 21.48$\pm$0.03 & -18.2 & -17.6 \\
WD93  & 135.587597 & -5.445684 & 24.08$\pm$0.10 & 23.76$\pm$0.07 & 25.67$\pm$0.66 & 23.04$\pm$0.07 & 24.12$\pm$0.39 & 27.1 & -0.1 \\
WD99  & 136.148961 & -5.454180 & 21.13$\pm$0.01 & 20.80$\pm$0.00 & 20.72$\pm$0.01 & 20.71$\pm$0.01 & 20.77$\pm$0.02 & 22.5 & -39.2 \\
WD100 & 136.088231 & -5.430176 & 23.97$\pm$0.08 & 23.14$\pm$0.04 & 22.71$\pm$0.05 & 22.30$\pm$0.04 & 22.52$\pm$0.10 & -21.9 & -5.5 \\
WD101 & 135.989002 & -4.168098 & 24.01$\pm$0.31 & 22.00$\pm$0.03 & 20.75$\pm$0.02 & 21.01$\pm$0.03 & 20.45$\pm$0.03 & -13.9 & 22.7 \\
WD102 & 135.818565 & -4.314596 & 25.22$\pm$0.35 & 24.11$\pm$0.09 & 23.53$\pm$0.09 & 23.77$\pm$0.12 & 23.92$\pm$0.31 & 21.0 & -16.1 \\
WD103 & 135.086246 & -4.525658 & 24.23$\pm$0.10 & 24.20$\pm$0.08 & 23.96$\pm$0.12 & 23.04$\pm$0.05 & 22.66$\pm$0.08 & 17.6 & 12.7 \\
WD105 & 135.863276 & -4.860796 & 23.17$\pm$0.03 & 23.07$\pm$0.03 & 22.83$\pm$0.04 & 22.49$\pm$0.03 & 22.18$\pm$0.06 & 19.0 & 8.0 \\
WD106 & 136.108813 & -4.887646 & 23.82$\pm$0.06 & 23.29$\pm$0.03 & 22.90$\pm$0.04 & 22.24$\pm$0.03 & 21.89$\pm$0.05 & -20.4 & 11.9 \\
WD107 & 136.298099 & -4.879626 & 23.30$\pm$0.04 & 23.48$\pm$0.04 & 23.27$\pm$0.07 & 23.10$\pm$0.07 & 23.18$\pm$0.18 & 10.3 & -18.0 \\
WD108 & 136.347112 & -4.857691 & 24.31$\pm$0.07 & 23.90$\pm$0.04 & 23.62$\pm$0.06 & 23.55$\pm$0.07 & 23.58$\pm$0.18 & -0.5 & -21.0 \\
WD109 & 136.191332 & -4.840274 & 24.23$\pm$0.10 & 24.29$\pm$0.10 & 23.97$\pm$0.10 & 23.37$\pm$0.09 & 23.65$\pm$0.26 & 18.2 & -9.2 \\
WD110 & 136.256013 & -5.020012 & 24.20$\pm$0.10 & 23.71$\pm$0.05 & 22.97$\pm$0.05 & 22.23$\pm$0.03 & 21.76$\pm$0.05 & -19.8 & 22.5 \\
WD111 & 136.212225 & -4.929040 & 23.76$\pm$0.06 & 23.23$\pm$0.03 & 22.81$\pm$0.05 & 22.84$\pm$0.05 & 22.92$\pm$0.12 & 17.2 & -23.1 \\
WD112 & 136.658616 & -5.005289 & 23.41$\pm$0.03 & 22.84$\pm$0.02 & 22.46$\pm$0.02 & 22.33$\pm$0.02 & 22.43$\pm$0.06 & -33.4 & 7.9 \\
\hline 
\end{tabular} 
\end{table*}

\begin{figure}
\vspace{0.7in}
\includegraphics[width=\columnwidth, bb = 36 237 577 455]{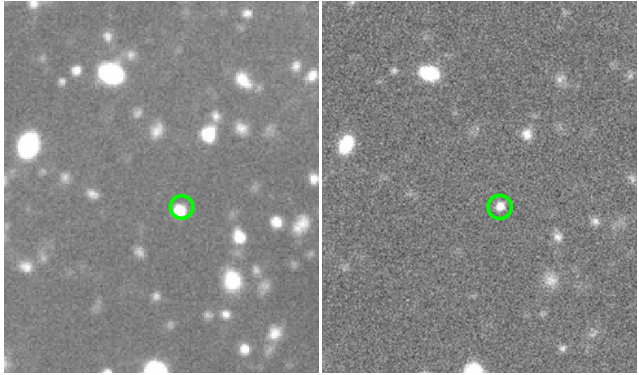}
\caption{One of the newly identified high proper motion white dwarfs in our program.
WD87, J090212.1-051642.7, has $g=23.7$ mag, and displays a proper motion of $95.3
\pm 5.1$ mas yr$^{-1}$. Our stacked DECam image is shown on the left panel,
whereas the original CFHTLS image is shown on the right. Green circles mark the
position of this object in the CFHTLS data.}
\label{fig:wd1355}
\end{figure}

Figure \ref{fig:wd1355} shows the field around one of the newly identified,
good white dwarf candidates in our program, WD87. This figure demonstrates
that our stacked DECam image is significantly deeper than the CFHT
Legacy Survey data. WD87, J090212.1-051642.7, is relatively
faint with $g=23.7$ mag, but its proper motion is clearly detected between
the original CFHT observations and our DECam data from 2014.

\subsection{Sample Size} 

There are 156,325 objects with $g=17.5-24.5$ mag in our field and we classify
59,517 of them as point sources based on their Flux Radius measurements in the
2.1-3.3 pixels range. We lose 14\% of the DECam area to trimming, and
an additional 9\% is lost when performing the 70 pixel cut for the distance of
the object from the edge. These cuts are necessary, partly due to the poor quality
of the stacked images near the edges and due to the worsening of the astrometric
calibration close to the edges. About 7.1\% of the remaining stars (hence $\sim$5.5\% of the total) in
CFHTLS were not detected on the DECam images. Of these, 52\% were too close to a
large bright object, 40\% were blended with a neighbour, and the remaining 8\%
were affected by image defects. Finally, 4.2\% of the remaining stars ($\sim$3\% of the original amount) were
eliminated due to their nearest neighbour being within $2 \arcsec$ in either the
CFHTLS or the DECam catalog. Therefore the total loss is $\sim$31.5\%.

The Besan\c{c}on Galaxy model \citep{Robi03} predicts $\sim$103
white dwarfs with $g<24.5$ mag and tangential velocities above 20 km s$^{-1}$ in our DECam field. 
Given the 31\% loss in our analysis, the expected number of high proper motion
white dwarfs is 71. This is similar to the number of white dwarf candidates
that we identify based on the reduced proper motion diagram. 

Tables 1 and 2 present the coordinates, $ugriz$ photometry from the CFHT
Legacy Survey, and proper motions for the remaining 78 good white dwarf candidates
and 33 MAYBEs. The good white dwarf candidates include three objects with
$\mu \approx 100$ mas yr$^{-1}$, WD69, WD73, and WD87, whereas the MAYBEs
include candidates with total proper motions up to 44 mas yr$^{-1}$.

\section{Sample Properties}

\subsection{Temperatures and Cooling Ages}

\begin{figure}
\includegraphics[width=\columnwidth, bb = 20 17 592 679]{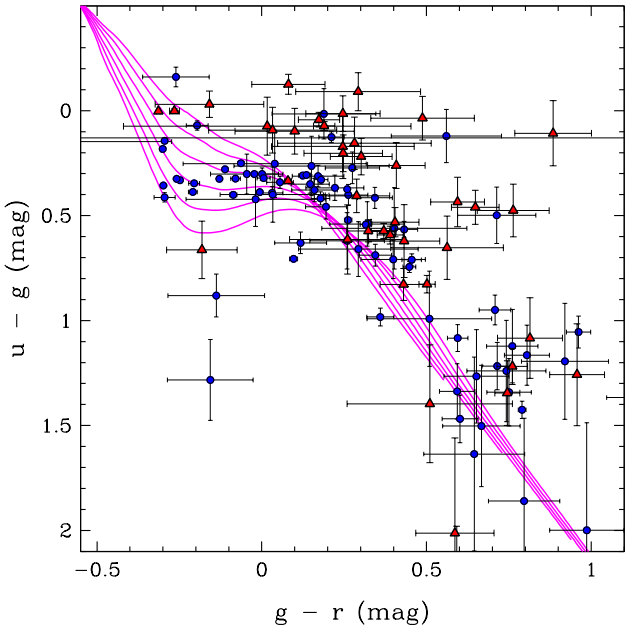}
\vspace{-0.7in}
\caption{A colour-colour diagram for our proper motion selected
sample of white dwarfs.  The colours for ``good'' (blue circles) and ``maybe'' (red triangles) white
dwarf candidates along with the predicted colours for pure hydrogen atmospheres
with $\log{g} = 7.0$ to 9.5 and $T_{\rm eff}$ from 120,000 to 1500 K are also
shown.}
\label{fig:colours}
\end{figure}

Figure \ref{fig:colours} displays the $u-g$ versus $g-r$ colour-colour diagram for
our white dwarf candidates along with the predicted colours for pure hydrogen
atmosphere white dwarfs. The colours for the majority of the good white dwarf
candidates are consistent with the models within the errors. On the other hand,
a significant fraction of the MAYBEs have redder colours than the models, which
could be due to contamination from M dwarf companions or our misidentification of
these targets as white dwarfs. Regardless of this issue, our white dwarf candidates
show a broad range of colours, indicating a broad range of temperatures, down to the
cool white dwarf regime.

We use pure hydrogen and pure helium atmosphere white dwarf models from
\citet{bergeron95} and \citet{Berg11} with the improvements discussed
in \citet{tremblay09} to fit the spectral energy distributions of our targets
to constrain their temperatures. Our model grid covers $T_{\rm eff} =$ 1500 to
45,000 K. 

Our fitting procedures are described in detail by \citet{Gian15}. The only difference
here is that due to lack of spectra and parallaxes for our targets, we
assume a surface gravity of $\log{g} = 8.0$. This is acceptable, given that the
main peak in the mass distribution of the white dwarfs in the Solar neighbourhood is
around 0.6 M$_{\odot}$ \citep{Trem11, giammichele12, limoges15}. Uncertainties arising from this assumption are further discussed in Section 4.2.

\begin{figure*}
\includegraphics[width=0.9\textwidth]{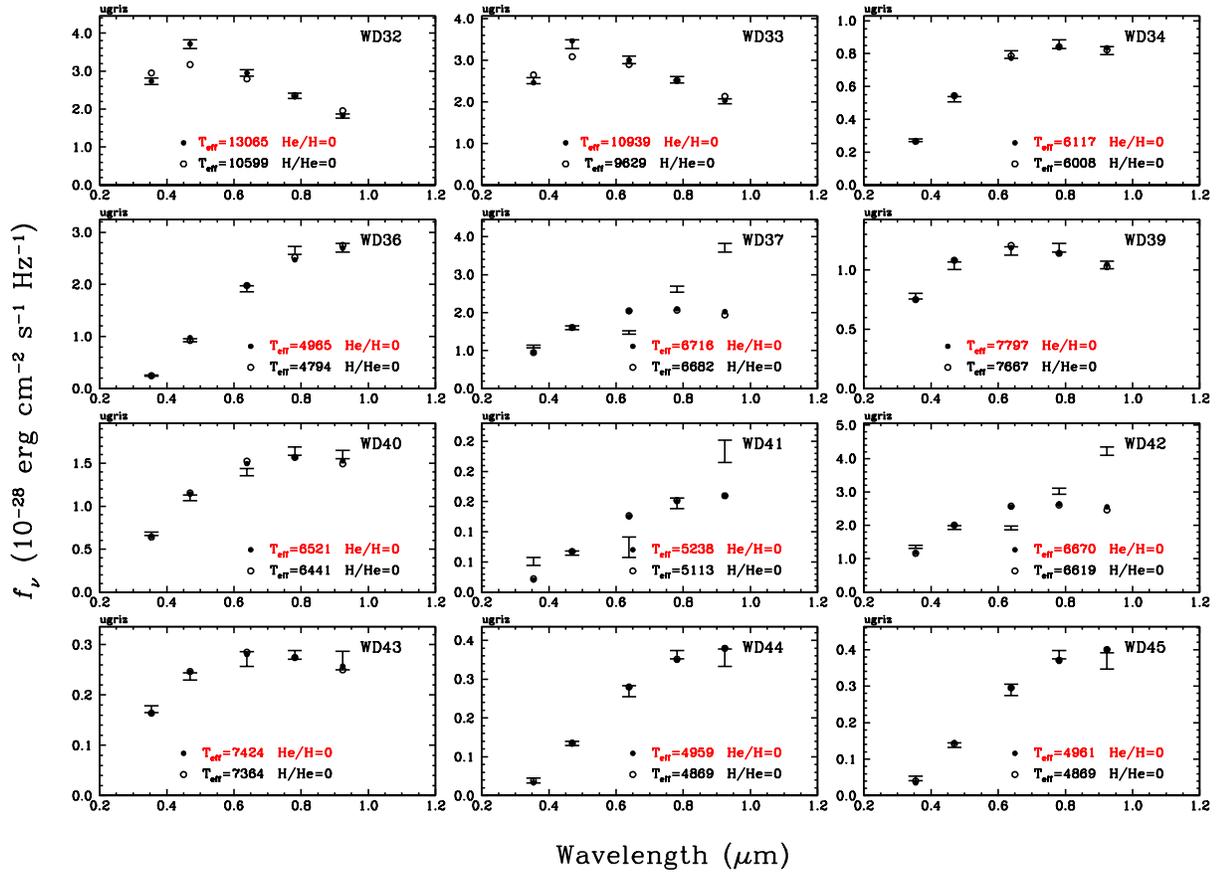}
\caption{Fits to the observed spectral energy distributions (error bars) of 12 of the
good white dwarf candidates with pure hydrogen (filled circles) and pure helium
atmosphere models (open circles).  The adopted atmospheric parameters are
emphasized in red.}
\label{fig:fit}
\end{figure*}

Figure \ref{fig:fit} displays representative spectral energy distributions and our pure H and pure He
model atmosphere fits to a dozen good white dwarf candidates in our sample. Given the
relatively faint magnitudes of our targets, only $ugriz$ photometry is available, which
limits the choice of composition, especially for cool white dwarfs. Hydrogen-rich white
dwarfs below about 5,000 K suffer from collision induced absorption due to molecular
hydrogen. Hence, the lack of significant infrared absorption could be a sign of a hydrogen
poor atmosphere.

The first two stars in Figure \ref{fig:fit}, WD32 and WD33, are warmer than about 10,000 K,
at which the differences between pure hydrogen and pure helium atmospheres are significant.
The spectral energy distributions of these two stars clearly indicate a hydrogen atmosphere
white dwarf. On the other hand, the rest of the stars in this figure are cooler than 8,000 K
where the differences between the hydrogen and helium atmosphere models in optical photometry
is minimal. Without near-infrared photometry, we assume pure hydrogen composition 
for our targets, unless the spectral energy distribution clearly favors the pure helium
atmosphere model. The latter is true for a few of our targets, namely WD68, WD84, WD97,
and one of the MAYBEs, WD4.  

All but three of the objects in Figure \ref{fig:fit} have spectral energy distributions that
are well matched by our models. However, the photometry for WD37, WD41, and WD42 is
too red to be explained by a single white dwarf. It is possible that these are white dwarf
+ M dwarf binary systems or they are contaminated by background sources. However,
all three of these targets look like point sources, and we detect $>$20 mas yr$^{-1}$
proper motions. In addition, one of these sources, WD42, is a variable object with
pulsation periods of 11-13 min, which demonstrate that it is a pulsating ZZ Ceti white
dwarf (see Section 5.2) likely in a white dwarf + M dwarf system \citep{pyrzas15}. Thus, we favour the
stellar binary explanation for these sources.

Tables 3 and 4 include temperature, distance, age, and tangential velocity estimates
for our white dwarf candidates. Our sample includes white dwarfs with $T_{\rm eff}=$
22,150 K down to 4250 K, which correspond to cooling ages of up to 8.4 Gyr for
$\log{g}=8$ white dwarfs. The estimated distances range from 147 pc to 4.9 kpc. 

\begin{table}
\centering
\scriptsize
\caption{Physical parameters of the good white dwarf candidates in the first field. Here and in Table 4
we assume a surface gravity of $\log{g}=8$.}
\begin{tabular}{lcrcr}
\hline
Target & $T_{\rm eff}$ & Distance  &  Age  & $V_{\rm tan}$ \\
Name  &    (K)  &   (pc)    & (Gyr) & (km s$^{-1}$)  \\
\hline
WD1 &   5960$\pm$220 &  853 & 2.28 &  89\\
WD2 &  16200$\pm$800 & 1776 & 0.15 & 184\\
WD5 &  15600$\pm$620 &  297 & 0.17 &  46\\
WD6 &  11140$\pm$430 &  641 & 0.45 & 103\\
WD7 &   4290$\pm$150 &  322 & 8.24 &  35\\
WD9 &   8260$\pm$300 &  758 & 0.99 &  80\\
WD10 &  6510$\pm$260 &  803 & 1.82 & 121\\
WD12 & 13040$\pm$660 &  842 & 0.30 & 103\\
WD14 &  7790$\pm$670 & 1368 & 1.15 & 155\\
WD15 &  6100$\pm$370 & 1005 & 2.15 &  97\\
WD16 &  4250$\pm$190 &  326 & 8.36 &  31\\
WD17 &  5390$\pm$80  &  342 & 3.63 &  36\\
WD18 &  5230$\pm$100 &  411 & 4.50 &  42\\
WD19 &  5020$\pm$230 &  597 & 5.61 & 163\\
WD20 & 13190$\pm$620 &  508 & 0.29 &  56\\
WD21 & 11970$\pm$680 & 1195 & 0.38 & 162\\
WD22 &  8920$\pm$310 &  531 & 0.81 &  61\\
WD23 &  7000$\pm$550 & 1231 & 1.52 & 118\\
WD24 &  6710$\pm$160 &  696 & 1.69 & 135\\
WD25 & 22150$\pm$2030& 3715 & 0.04 & 398\\
WD28 &  7140$\pm$160 &  534 & 1.45 &  79\\
WD29 &  5090$\pm$90  &  365 & 5.24 &  45\\
WD30 &  4600$\pm$110 &  352 & 7.25 &  41\\
WD31 &  7980$\pm$240 &  664 & 1.08 &  67\\
WD32 & 13070$\pm$640 &  502 & 0.30 &  53\\
WD33 & 10940$\pm$410 &  432 & 0.48 &  47\\
WD34 &  6120$\pm$70 &   357 & 2.13 &  75\\
WD36 &  4970$\pm$50 &   147 & 5.86 &  30\\
WD37 &  6720$\pm$130 &  256 & 1.68 &  28\\
WD39 &  7800$\pm$210 &  423 & 1.15 &  76\\
WD40 &  6520$\pm$100 &  285 & 1.81 &  31\\
WD41 &  5240$\pm$200 &  658 & 4.46 &  66\\
WD42 &  6670$\pm$120 &  226 & 1.71 &  29\\
WD43 &  7430$\pm$230 &  807 & 1.31 &  92\\
WD44 &  4960$\pm$120 &  391 & 5.88 &  70\\
WD45 &  4960$\pm$120 &  381 & 5.87 &  63\\
WD46 &  8000$\pm$220 &  434 & 1.07 &  85\\
WD47 &  7510$\pm$310 &  959 & 1.27 & 106\\
WD48 &  5950$\pm$60 &   334 & 2.29 &  52\\
WD49 &  5260$\pm$70 &   302 & 4.35 &  45\\
WD50 &  7000$\pm$220 &  766 & 1.52 &  73\\
WD51 &  6540$\pm$380 & 1055 & 1.80 & 103\\
WD52 &  8070$\pm$260 &  759 & 1.05 &  97\\
WD53 &  7160$\pm$260 &  844 & 1.43 &  87\\
WD54 &  5830$\pm$140 &  564 & 2.43 &  59\\
WD55 &  4940$\pm$90 &   357 & 5.98 &  38\\
WD56 &  5130$\pm$200 &  619 & 5.06 &  66\\
WD59 &  6380$\pm$90 &   285 & 1.91 &  94\\
WD61 &  4700$\pm$80 &   268 & 6.92 &  34\\
WD62 &  6870$\pm$120 &  205 & 1.59 &  25\\
WD63 &  7470$\pm$690 & 1450 & 1.29 & 138\\
WD64 &  6710$\pm$180 &  721 & 1.69 & 129\\
WD65 &  5260$\pm$210 &  602 & 4.36 &  62\\
WD66 &  6070$\pm$170 &  710 & 2.17 & 208\\
WD67 &  7600$\pm$350 &  954 & 1.23 & 136\\
WD68 &  9390$\pm$240 &  388 & 0.78 &  58\\
WD69 & 13170$\pm$630 &  601 & 0.29 & 290\\
WD73 & 15370$\pm$620 &  261 & 0.18 & 147\\
WD74 &  5280$\pm$190 &  626 & 4.24 &  74\\
WD75 &  6100$\pm$100 &  550 & 2.15 &  66\\
WD76 &  6190$\pm$90 &   514 & 2.07 &  76\\
WD77 &  8050$\pm$1000& 1692 & 1.06 & 195\\
WD78 &  6550$\pm$100 &  318 & 1.79 &  36\\
WD79 &  6570$\pm$120 &  511 & 1.78 &  52\\
WD80 &  6380$\pm$330 & 1049 & 1.92 & 134\\
WD81 &  4850$\pm$220 &  543 & 6.38 & 132\\
WD83 &  7030$\pm$260 &  893 & 1.50 &  97\\
WD84 &  9270$\pm$230 &  474 & 0.80 &  80\\
WD86 &  5380$\pm$260 &  741 & 3.72 & 163\\
WD87 & 12500$\pm$2940& 2855 & 0.33 & 1295\\
WD88 &  4750$\pm$190 &  494 & 6.75 &  81\\
WD89 &  4860$\pm$200 &  521 & 6.31 &  55\\
WD90 &  7530$\pm$190 &  471 & 1.26 &  57\\
WD94 &  6460$\pm$90 &   360 & 1.86 &  43\\
WD95 &  5220$\pm$90 &   389 & 4.59 &  84\\
WD96 &  8430$\pm$270 &  586 & 0.94 & 130\\
WD97 &  8510$\pm$210 &  525 & 1.01 &  86\\
WD98 &  8280$\pm$1360& 1764 & 0.98 & 192\\
\hline
\end{tabular}
\end{table}

\begin{table}
\centering
\scriptsize
\caption{Physical parameters of the ``maybe'' white dwarf candidates in the first field.}
\begin{tabular}{lcrcr}
\hline
Target & $T_{\rm eff}$ & Distance  &  Age  & $V_{\rm tan}$ \\
Name  &    (K)  &   (pc)    & (Gyr) & (km s$^{-1}$)  \\
\hline
WD3   &  5930$\pm$80  &  448 & 2.32 &  50 \\
WD4   & 12570$\pm$520 & 1019 & 0.36 & 106 \\
WD8   &  5540$\pm$200 &  637 & 3.06 &  65 \\
WD11  &  6650$\pm$280 &  940 & 1.73 &  96 \\
WD13  &  5720$\pm$270 &  748 & 2.62 &  85 \\
WD26  &  6560$\pm$460 & 1176 & 1.79 & 126 \\
WD27  &  5910$\pm$260 &  857 & 2.34 &  89 \\
WD35  & 17120$\pm$3870& 4923 & 0.13 & 513 \\
WD38  &  6540$\pm$340 &  958 & 1.80 & 126 \\
WD57  &  4340$\pm$160 &  362 & 8.09 &  35 \\
WD58  &  7180$\pm$510 & 1167 & 1.42 & 172 \\
WD60  &  5250$\pm$220 &  685 & 4.41 &  67 \\
WD70  &  7320$\pm$550 & 1193 & 1.36 & 141 \\
WD71  &  5160$\pm$170 &  578 & 4.88 &  60 \\
WD72  & 17620$\pm$750 &  627 & 0.11 &  93 \\
WD82  &  6360$\pm$100 &  543 & 1.93 &  89 \\
WD85  &  7460$\pm$580 & 1347 & 1.29 & 141 \\
WD91  &  6890$\pm$470 & 1204 & 1.58 & 124 \\
WD92  &  5130$\pm$80  &  375 & 5.04 &  40 \\
WD93  &  6810$\pm$140 &  509 & 1.63 &  61 \\
WD99  &  7300$\pm$440 & 1048 & 1.37 & 135 \\
WD100 &  7750$\pm$200 &  330 & 1.17 &  71 \\
WD101 &  5120$\pm$100 &  408 & 5.07 &  52 \\
WD102 &  6910$\pm$520 & 1283 & 1.57 & 161 \\
WD103 &  5010$\pm$180 &  516 & 5.63 &  53 \\
WD105 &  6930$\pm$170 &  725 & 1.56 &  71 \\
WD106 &  5530$\pm$90  &  457 & 3.07 &  51 \\
WD107 &  8370$\pm$600 & 1211 & 0.96 & 119 \\
WD108 &  6870$\pm$340 & 1056 & 1.59 & 106 \\
WD109 &  7040$\pm$540 & 1202 & 1.50 & 117 \\
WD110 &  6350$\pm$110 &  577 & 1.94 &  60 \\
WD111 &  4610$\pm$110 &  333 & 7.22 &  48 \\
WD112 &  6490$\pm$190 &  696 & 1.83 &  95 \\
\hline
\end{tabular}
\end{table}

\subsection{Kinematic Properties: Halo versus Disk}

\begin{figure}
\includegraphics[width=2.4in,angle=-90]{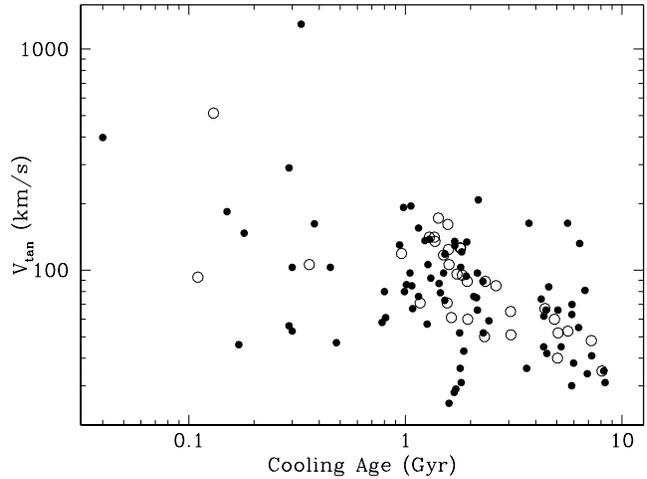}
\caption{Tangential velocity versus age for the good (filled) and probable (open
circles) white dwarf candidates in the first field. Both $V_{\rm tan}$ and the
cooling age strongly depend on the mass of the white dwarf, which is assumed to
be $M\approx0.6 M_{\odot}$ (i.e., $\log{g}=8$) in this analysis.}
\label{fig:vtan}
\end{figure}

Figure \ref{fig:vtan} presents tangential velocities and cooling ages for our
high proper motion white dwarf candidates. There are two candidates, WD87 and
WD35 (which is classified as a MAYBE), with tangential velocities larger than
the escape speed from the Galaxy. 

Both velocities and cooling ages
strongly depend on the choice of mass, or surface gravity for our targets.
The white dwarf mass distribution peaks at about 0.6 $M_{\odot}$ and there is
another peak at lower masses (0.4 $M_{\odot}$) and a tail toward higher masses
\citep{tremblay13}. Hence, our choice of $\log{g}=8$ is appropriate for most targets,
and for studying the ensemble properties, but the cooling ages and especially tangential
velocities may be skewed toward higher values if the surface gravity is significantly
different. For example, at $T_{\rm eff}$ = 12,000K, increasing the assumed surface gravity
by 1 dex lowers the distance estimate and thus the tangential velocity by a factor of two.
Hence, WD87 and WD35 are likely more massive than the canonical 0.6 $M_{\odot}$, as it is
highly unlikely to find two such objects in our relatively small sample of white dwarfs.

Figure \ref{fig:vtan}
surprisingly shows a trend in which younger objects tend to have higher
tangential velocities. This is highly unexpected, as halo
objects (therefore older and cooler white dwarfs) should have larger tangential
velocities. Particularly, WD25 has a tangential velocity of 400 km s$^{-1}$, and
a cooling age of only 40 million years. It is possible that WD25 could be a newly
born halo white dwarf with $M \approx 0.53 M_{\odot}$ \citep{bergeron01}. However,
this would be a rare occurrence, since most of the halo population is old.
The Besan\c{c}on Galaxy model \citep{Robi03} predicts 0-1 halo white dwarfs in our field
down to $g=24.5$ mag. Some of the fastest moving objects could in fact still belong to the
Galactic disc, but have a $\log{g} > 8.0$.

Given the significant errors on the estimated distances, the tangential velocity distribution is predicted to range between 19 and 110 km s$^{-1}$ for the thin disk and between 53 and 133 km s$^{-1}$ for the thick disk populations. Moreover, the thick disk white dwarfs
are predicted to be outnumbered by the thin disk sample by a factor of 15. 
The majority of the objects in our sample have tangential
velocities $< 200$ km s$^{-1}$, thus are consistent with Galactic disk membership.

\section{The Light Curves}

\subsection{Building the Light Curves}

In order to generate lightcurves for all of the white dwarf targets, we used the IRAF
$phot$ package to perform aperture photometry on all of the images collected
with DECam. Since we are interested in identifying variable objects, only relative
photometry is needed for our program, and we do not perform absolute photometric
calibration of our DECam observations. 

To account for short-term changes in the atmosphere (such as cloud
coverage or haze), as well as the change in airmass, we use six bright,
unsaturated, non-variable stars in each CCD as reference stars. 
Given the large field of view of DECam, the image quality differs for different CCDs.
Hence, the selection of reference stars and our calibration procedures are done
separately for each CCD. After shifting
the six reference stars to the same magnitude scale, we use a sigma-clipping algorithm
to reject bad points that are affected by cosmic rays or CCD defects, and
we take their weighted mean to create a reference light curve for each CCD.
To calibrate the relative photometry for our white dwarf targets, we subtract the
appropriate calibration light curve, given the CCD that includes each target.
Note that this process was run separately for each night, and for each CCD that
contained one or more targets.

The reference stars chosen to calibrate the light curves are typically redder
than our white dwarf targets. Hence, airmass related effects are still present
in the light curves of many of our targets and they lead to significant peaks
especially at 4 cycles per day in the Fourier Transforms. We fitted third degree polynomials
to the target light curves to remove this effect.

\begin{figure*}
\includegraphics[width=\textwidth]{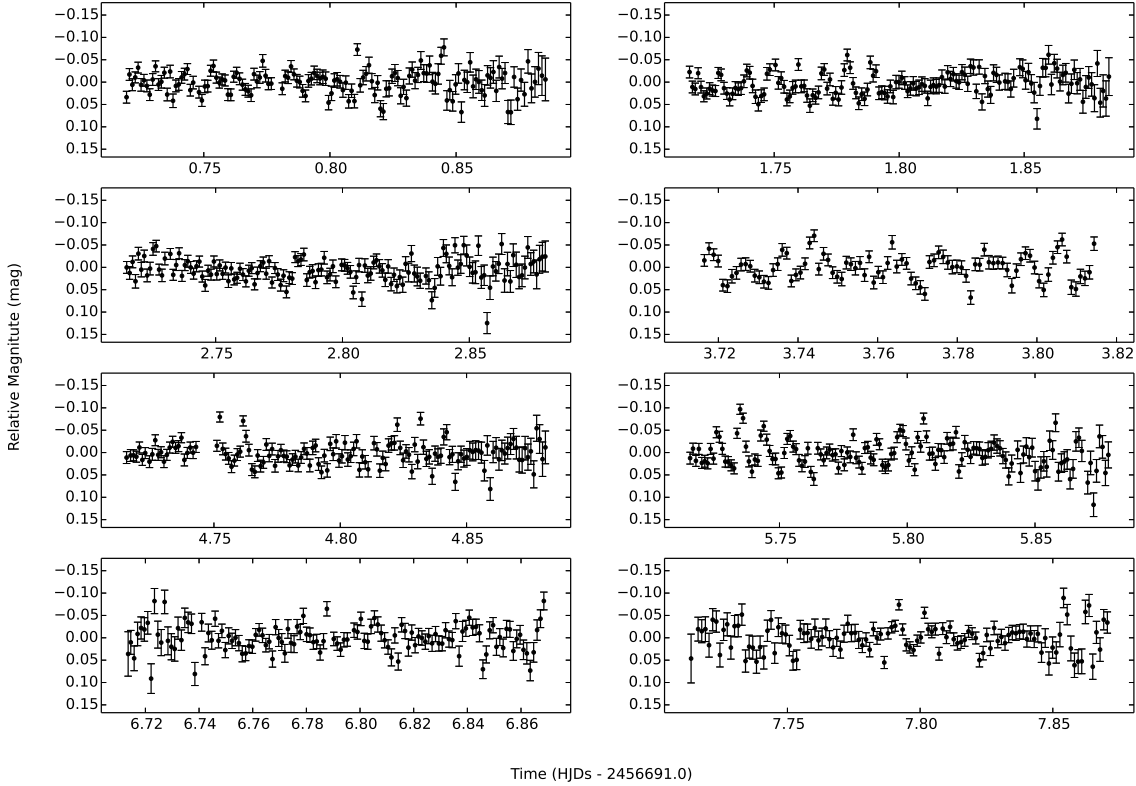}
\caption{DECam time-series photometry of WD42 over 8 half-nights. The lightcurve
shows significant sinusoidal variations, especially visible in nights two and
four, making this one of the faintest ZZ Ceti white dwarfs currently known.}
\label{fig:WD42}
\end{figure*}

\begin{figure}
\includegraphics[width=\columnwidth]{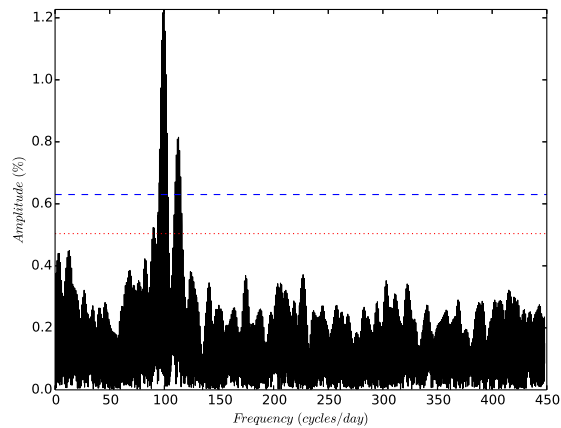}
\caption{Fourier transform of the WD42 light curve. The red dotted line and the
blue dashed line show the $4<$A$>$ and $5<$A$>$ detection limits, respectively.
There are two significant peaks, indicating pulsation frequencies of 99.6 and
112.9 cycles day$^{-1}$.}
\label{fig:FT42}
\end{figure}

\subsection{A New Variable System}

We use the Period04\footnote{https://www.univie.ac.at/tops/Period04/} package
to calculate Discrete Fourier Transforms for each of our white dwarf targets.
We compare the amplitudes of the observed peaks with the median amplitude ($<$A$>$)
in the Fourier Transform, and only consider the peaks above $5<$A$>$ as significant.
Some of our targets still show significant peaks at 4, 6, 8, or 12 cycles per day due
to our observing window. However, there is only one target that is clearly variable.

Figure \ref{fig:WD42} shows the 8-night-long light curve of WD42. Sinusiodal variations
in each night are clearly visible in this light curve. Figure \ref{fig:FT42} shows
the Fourier Transform of this light curve. There are two significant peaks at
$99.5889 \pm 0.0069$ and $112.9466 \pm 1.8662$ cycles per day with amplitudes
of $1.22 \pm 0.12$\% and $0.79 \pm 0.25$\%, respectively. The observed range of
periods, 765-868 s, are consistent with the pulsation periods seen in ZZ Ceti
white dwarfs \citep[e.g.,][]{mukadam04}. 

ZZ Ceti white dwarfs are found in a narrow instability strip around
$T_{\rm eff}$ $\approx$ 12,000 K \citep{gianninas11}. However, our model fits to the
spectral energy distribution of WD42 (see Fig. \ref{fig:fit}) indicate an effective temperature
of 6670 K for a single white dwarf. These models clearly fail to reproduce 
the spectral energy distribution of WD42. A likely explanation for the observed
colours of WD42 is that it is a pulsating white dwarf + M dwarf system in which
the white dwarf dominates the photometry in the blue ($ug$ bands) and the M dwarf
dominates in the red. With an apparent magnitude of $g=$ 20.64, WD42 becomes
the second faintest pulsating white dwarf known after the white dwarf companion
of PSR J1738+0333 \citep{Kili15}.

\subsection{No Transits}

Solid body transits around white dwarfs last 1-2 min \citep[e.g.][]{brown11}.
Hence, such events would affect only one or two photometric points for each
orbital period. After constructing 8-night-long lightcurves for
all of the white dwarf candidates and MAYBEs, we checked all of them for the presence of
significant ($\ge 4 \sigma$) dips, and we visually inspected the images with
potential transits. One of our targets, WD46, showed significant photometric
dips at the end of the night, for several nights. A careful inspection of the images
with the dips show that WD46 is close to the edge of the chip and the point spread
function is rather elliptical at its location in these images. Hence, the observed
photometric dips are not real. None of the remaining targets show any
significant dips that could be attributed to an eclipsing planet orbiting
around its host star.

Given our 4 h long observing window each night, we are sensitive to
100\% of transiting objects with orbital periods of 4 h or less, assuming they
cause significant eclipses. However, we expect the detection rate to fall significantly
at longer orbital periods, and especially at 6, 8, 12, 16, 24 and 28 h, which are
discrete frequencies of our observing window of 4 h per night.

To estimate our transit detection efficiency, we simulate an 8-day-long set of
lightcurves, each with cadences of 90 seconds, presenting eclipses the length
of one data point at periods ranging from 2 to 30 hours, with increments of 30 minutes.
We then filter the light curves through our observing window and shift the times of eclipses
to probe all possible configurations. We require at least two eclipse
events to call it a detection.

\begin{figure}
\includegraphics[width=\columnwidth]{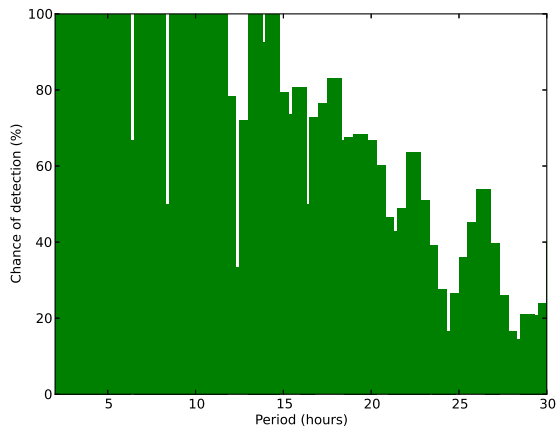}
\caption{The probability of transit detection (green area) as a function of the
orbital period based on our cadence and observing window of 8 half-nights. The
range of periods covers the extent of the Habitable Zone of a 0.6 M$_{\odot}$
white dwarf. The cumulative detection rate for $P\le30$ h is 68.5\%.}
\label{fig:detection}
\end{figure}

Figure \ref{fig:detection} presents the transit detection probability for a flat distribution of periods. We see
an overall trend of decreasing probability as the orbital period increases.
The total shaded area in the graph represents the cumulative
probability of detecting a transiting object in the sample, which
corresponds to 68.5\%. In other words, our DECam observations are
capable of detecting 68.5\% of the significant transits with periods less than 30 h.
The detectability of a given planet depends on the magnitude of its host star.
Dimmer stars show higher scatter in their light curves, and this limits the detection of asteroids and moons
around the fainter targets. Table 5 shows the minimum  depth (in percentage) required for eclipses to be
detected at the $4\sigma$ level when probing stars of different magnitudes. Assuming that the majority
of our targets are average mass white dwarfs with a radius comparable to Earth, our observations are
sensitive to transits by moon-sized objects for targets brighter than $g=20$ mag, and by Earth-sized objects
for all targets.

\begin{table}
\centering
\scriptsize
\caption{Minimum detectable transit depth for a variety of magnitudes present in our sample}
\begin{tabular}{lcc}
\hline
Target & Magnitude & Minimum Transit Depth  \\
Name  &    (mag)  &   (\%)      \\
\hline
WD73  & 18.22 & 2.7 \\
WD33  & 20.04 & 7.5 \\
WD12  & 21.09 &  11.1 \\
WD34  & 22.10 &  36.7 \\
WD79  & 22.58 &  51.0 \\
WD50  & 23.01 &  63.2 \\
WD23  & 24.03 &  86.4 \\
WD81  & 24.50 &  92.6 \\
\hline
\end{tabular}
\end{table}

The probability of an eclipse is 1\% for an Earth-like
planet orbiting in the habitable zone of a white dwarf, which typically extends between 0.005 and 0.02 AU for a 0.6 M$_{\odot}$ white dwarf \citep{Agol11}. Therefore, our cumulative
detection rate corresponds to an expected detection rate of 0.7\% due to our
observing window from the ground. Hence, we would expect to find 0.8 planets in
a sample of 111 stars, if each white dwarf had an earth-mass planetary companion within
its habitable zone. Therefore, the lack of detection of eclipses in our sample
of 111 white dwarf candidates is not surprising.

\section{Conclusions}

We present the results from the first minute-cadence survey of a large number of
white dwarf candidates observed with DECam. We identify 111 high proper motion white dwarf candidates
brighter than $g=24.5$ mag in a single DECam pointing. We estimate temperatures, cooling
ages, and tangential velocities for each object and demonstrate that our targets
are consistent with thin and thick disk white dwarfs.

We create light curves
for each white dwarf, spanning 8 half-nights. We identify a $g=20.64$ mag pulsating
ZZ Ceti white dwarf, most likely in a binary system with an M dwarf companion. 
We do not find any eclipsing systems in this first field, but given the probabiliy
of eclipses of 1\% and our observing window from the ground, this is not surprising.
However, this work demonstrates the feasibilty of using DECam to search for minute-cadence
transits around white dwarfs. In addition to the high proper motion white dwarfs,
the Besan\c{c}on Galaxy model predicts 400 other white dwarfs with $\mu<$20 mas yr$^{-1}$
in one of our DECam fields. Image subtraction routines can be used to search for variability
for all of the sources in our DECam field, including the non-moving white dwarfs. Such
a study with High Order Transform of PSF and Template Subtraction \citep[HOTPANTS,][]{becker15}
is currently underway and it will be presented in a future publication. 
Given the probability of 0.7\% of finding a transit around a white dwarf,
increasing the size of the white dwarf sample to several hundreds would
enable us to find the first solid-body planetary companion, if such systems exist.

\section*{Acknowledgements} We gratefully acknowledge the support of the NSF and
NASA under grants AST-1312678 and NNX14AF65G, respectively. Based on observations at Cerro Tololo Inter-American Observatory, National Optical Astronomy Observatory (NOAO Prop. ID: 2014A-0073; PI: M. Kilic), which is operated by the Association of Universities for Research in Astronomy (AURA) under a cooperative agreement with the National Science Foundation. This project used
data obtained with the Dark Energy Camera (DECam), which was constructed by the
Dark Energy Survey (DES) collaboration.  Funding for the DES Projects has been
provided by the U.S. Department of Energy, the U.S. National Science Foundation,
the Ministry of Science and Education of Spain, the Science and Technology
Facilities Council of the United Kingdom, the Higher Education Funding Council
for England, the National Center for Supercomputing Applications at the
University of Illinois at Urbana-Champaign, the Kavli Institute of Cosmological
Physics at the University of Chicago, the Center for Cosmology and
Astro-Particle Physics at the Ohio State University, the Mitchell Institute for
Fundamental Physics and Astronomy at Texas A\&M University, Financiadora de
Estudos e Projetos, Funda{\c c}{\~a}o Carlos Chagas Filho de Amparo {\`a}
Pesquisa do Estado do Rio de Janeiro, Conselho Nacional de Desenvolvimento
Cient{\'i}fico e Tecnol{\'o}gico and the Minist{\'e}rio da Ci{\^e}ncia,
Tecnologia e Inovac{\~a}o, the Deutsche Forschungsgemeinschaft, and the
Collaborating Institutions in the Dark Energy Survey.  The Collaborating
Institutions are Argonne National Laboratory, the University of California at
Santa Cruz, the University of Cambridge, Centro de Investigaciones
En{\'e}rgeticas, Medioambientales y Tecnol{\'o}gicas-Madrid, the University of
Chicago, University College London, the DES-Brazil Consortium, the University of
Edinburgh, the Eidgen{\"o}ssische Technische Hoch\-schule (ETH) Z{\"u}rich,
Fermi National Accelerator Laboratory, the University of Illinois at
Urbana-Champaign, the Institut de Ci{\`e}ncies de l'Espai (IEEC/CSIC), the
Institut de F{\'i}sica d'Altes Energies, Lawrence Berkeley National Laboratory,
the Ludwig-Maximilians Universit{\"a}t M{\"u}nchen and the associated Excellence
Cluster Universe, the University of Michigan, {the} National Optical Astronomy
Observatory, the University of Nottingham, the Ohio State University, the
University of Pennsylvania, the University of Portsmouth, SLAC National
Accelerator Laboratory, Stanford University, the University of Sussex, and Texas
A\&M University.\\

\bibliographystyle{mn2e} 
\bibliography{biblio}

\begin{thebibliography}{37}
\expandafter\ifx\csname natexlab\endcsname\relax\def\natexlab#1{#1}\fi

\bibitem[{{Agol}(2011)}]{Agol11}
{Agol} E., 2011, \apjl, 731, L31

\bibitem[{{Alcock} {et~al}\mbox{.}(2000){Alcock}, {Allsman}, {Alves},
  {Axelrod}, {Becker}, {Bennett}, {Cook}, {Dalal}, {Drake}, {Freeman}, {Geha},
  {Griest}, {Lehner}, {Marshall}, {Minniti}, {Nelson}, {Peterson}, {Popowski},
  {Pratt}, {Quinn}, {Stubbs}, {Sutherland}, {Tomaney}, {Vandehei}, \&
  {Welch}}]{alcock00}
{Alcock} C. {et~al.}, 2000, \apj, 542, 281

\bibitem[{{Becker}(2015)}]{becker15}
{Becker} A., 2015, {HOTPANTS: High Order Transform of PSF ANd Template
  Subtraction}. Astrophysics Source Code Library

\bibitem[{{Bergeron}, {Leggett} \& {Ruiz}(2001){Bergeron}, {Leggett}, \&
  {Ruiz}}]{bergeron01}
{Bergeron} P., {Leggett} S.~K., {Ruiz} M.~T., 2001, \apjs, 133, 413

\bibitem[{{Bergeron}, {Wesemael} \& {Beauchamp}(1995){Bergeron}, {Wesemael}, \&
  {Beauchamp}}]{bergeron95}
{Bergeron} P., {Wesemael} F., {Beauchamp} A., 1995, \pasp, 107, 1047

\bibitem[{{Bergeron} {et~al}\mbox{.}(2011){Bergeron}, {Wesemael}, {Dufour},
  {Beauchamp}, {Hunter}, {Saffer}, {Gianninas}, {Ruiz}, {Limoges}, {Dufour},
  {Fontaine}, \& {Liebert}}]{Berg11}
{Bergeron} P. {et~al.}, 2011, \apj, 737, 28

\bibitem[{{Bernstein} {et~al}\mbox{.}(2012){Bernstein}, {Kessler}, {Kuhlmann},
  {Biswas}, {Kovacs}, {Aldering}, {Crane}, {D'Andrea}, {Finley}, {Frieman},
  {Hufford}, {Jarvis}, {Kim}, {Marriner}, {Mukherjee}, {Nichol}, {Nugent},
  {Parkinson}, {Reis}, {Sako}, {Spinka}, \& {Sullivan}}]{bernstein12}
{Bernstein} J.~P. {et~al.}, 2012, \apj, 753, 152

\bibitem[{{Bertin} \& {Arnouts}(1996)}]{bertin96}
{Bertin} E., {Arnouts} S., 1996, \aaps, 117, 393

\bibitem[{{Brown} {et~al}\mbox{.}(2011){Brown}, {Kilic}, {Hermes}, {Allende
  Prieto}, {Kenyon}, \& {Winget}}]{brown11}
{Brown} W.~R., {Kilic} M., {Hermes} J.~J., {Allende Prieto} C., {Kenyon} S.~J.,
  {Winget} D.~E., 2011, \apjl, 737, L23

\bibitem[{{Debes}, {Walsh} \& {Stark}(2012){Debes}, {Walsh}, \&
  {Stark}}]{debes12}
{Debes} J.~H., {Walsh} K.~J., {Stark} C., 2012, \apj, 747, 148

\bibitem[{{Drake} {et~al}\mbox{.}(2009){Drake}, {Djorgovski}, {Mahabal},
  {Beshore}, {Larson}, {Graham}, {Williams}, {Christensen}, {Catelan},
  {Boattini}, {Gibbs}, {Hill}, \& {Kowalski}}]{drake09}
{Drake} A.~J. {et~al.}, 2009, \apj, 696, 870

\bibitem[{{Flaugher}(2005)}]{flaugher05}
{Flaugher} B., 2005, International Journal of Modern Physics A, 20, 3121

\bibitem[{{G{\"a}nsicke} {et~al}\mbox{.}(2016){G{\"a}nsicke}, {Aungwerojwit},
  {Marsh}, {Dhillon}, {Sahman}, {Veras}, {Farihi}, {Chote}, {Ashley},
  {Arjyotha}, {Rattanasoon}, {Littlefair}, {Pollacco}, \&
  {Burleigh}}]{gansike16}
{G{\"a}nsicke} B.~T. {et~al.}, 2016, \apjl, 818, L7

\bibitem[{{Giammichele}, {Bergeron} \& {Dufour}(2012){Giammichele}, {Bergeron},
  \& {Dufour}}]{giammichele12}
{Giammichele} N., {Bergeron} P., {Dufour} P., 2012, \apjs, 199, 29

\bibitem[{{Gianninas}, {Bergeron} \& {Ruiz}(2011){Gianninas}, {Bergeron}, \&
  {Ruiz}}]{gianninas11}
{Gianninas} A., {Bergeron} P., {Ruiz} M.~T., 2011, \apj, 743, 138

\bibitem[{{Gianninas} {et~al}\mbox{.}(2015){Gianninas}, {Curd}, {Thorstensen},
  {Kilic}, {Bergeron}, {Andrews}, {Canton}, \& {Ag{\"u}eros}}]{Gian15}
{Gianninas} A., {Curd} B., {Thorstensen} J.~R., {Kilic} M., {Bergeron} P.,
  {Andrews} J.~J., {Canton} P., {Ag{\"u}eros} M.~A., 2015, \mnras, 449, 3966

\bibitem[{{Ivezi{\'c}} {et~al}\mbox{.}(2007){Ivezi{\'c}}, {Smith}, {Miknaitis},
  {Lin}, {Tucker}, {Lupton}, {Gunn}, {Knapp}, {Strauss}, {Sesar}, {Doi},
  {Tanaka}, {Fukugita}, {Holtzman}, {Kent}, {Yanny}, {Schlegel}, {Finkbeiner},
  {Padmanabhan}, {Rockosi}, {Juri{\'c}}, {Bond}, {Lee}, {Stoughton}, {Jester},
  {Harris}, {Harding}, {Morrison}, {Brinkmann}, {Schneider}, \&
  {York}}]{ivezic07}
{Ivezi{\'c}} {\v Z}. {et~al.}, 2007, \aj, 134, 973

\bibitem[{{Jura}(2003)}]{Jura03}
{Jura} M., 2003, \apjl, 584, L91

\bibitem[{{Kaiser} {et~al}\mbox{.}(2010){Kaiser}, {Burgett}, {Chambers},
  {Denneau}, {Heasley}, {Jedicke}, {Magnier}, {Morgan}, {Onaka}, \&
  {Tonry}}]{kaiser10}
{Kaiser} N. {et~al.}, 2010, in Society of Photo-Optical Instrumentation
  Engineers (SPIE) Conference Series, Vol. 7733, Society of Photo-Optical
  Instrumentation Engineers (SPIE) Conference Series, p.~0

\bibitem[{{Kilic} {et~al}\mbox{.}(2015){Kilic}, {Hermes}, {Gianninas}, \&
  {Brown}}]{Kili15}
{Kilic} M., {Hermes} J.~J., {Gianninas} A., {Brown} W.~R., 2015, \mnras, 446,
  L26

\bibitem[{{Kilic} {et~al}\mbox{.}(2010){Kilic}, {Leggett}, {Tremblay}, {von
  Hippel}, {Bergeron}, {Harris}, {Munn}, {Williams}, {Gates}, \&
  {Farihi}}]{kilic10}
{Kilic} M. {et~al.}, 2010, \apjs, 190, 77

\bibitem[{{Kilic} {et~al}\mbox{.}(2006){Kilic}, {Munn}, {Harris}, {Liebert},
  {von Hippel}, {Williams}, {Metcalfe}, {Winget}, \& {Levine}}]{kilic06}
{Kilic} M. {et~al.}, 2006, \aj, 131, 582

\bibitem[{{Limoges}, {Bergeron} \& {L{\'e}pine}(2015){Limoges}, {Bergeron}, \&
  {L{\'e}pine}}]{limoges15}
{Limoges} M.-M., {Bergeron} P., {L{\'e}pine} S., 2015, \apjs, 219, 19

\bibitem[{{Monet} {et~al}\mbox{.}(2003){Monet}, {Levine}, {Canzian}, {Ables},
  {Bird}, {Dahn}, {Guetter}, {Harris}, {Henden}, {Leggett}, {Levison},
  {Luginbuhl}, {Martini}, {Monet}, {Munn}, {Pier}, {Rhodes}, {Riepe}, {Sell},
  {Stone}, {Vrba}, {Walker}, {Westerhout}, {Brucato}, {Reid}, {Schoening},
  {Hartley}, {Read}, \& {Tritton}}]{Mone03}
{Monet} D.~G. {et~al.}, 2003, \aj, 125, 984

\bibitem[{{Mukadam} {et~al}\mbox{.}(2004){Mukadam}, {Winget}, {von Hippel},
  {Montgomery}, {Kepler}, \& {Costa}}]{mukadam04}
{Mukadam} A.~S., {Winget} D.~E., {von Hippel} T., {Montgomery} M.~H., {Kepler}
  S.~O., {Costa} A.~F.~M., 2004, \apj, 612, 1052

\bibitem[{{Pyrzas} {et~al}\mbox{.}(2015){Pyrzas}, {G{\"a}nsicke}, {Hermes},
  {Copperwheat}, {Rebassa-Mansergas}, {Dhillon}, {Littlefair}, {Marsh},
  {Parsons}, {Savoury}, {Schreiber}, {Barros}, {Bento}, {Breedt}, \&
  {Kerry}}]{pyrzas15}
{Pyrzas} S. {et~al.}, 2015, \mnras, 447, 691

\bibitem[{{Rappaport} {et~al}\mbox{.}(2016){Rappaport}, {Gary}, {Kaye},
  {Vanderburg}, {Croll}, {Benni}, \& {Foote}}]{rappaport16}
{Rappaport} S., {Gary} B.~L., {Kaye} T., {Vanderburg} A., {Croll} B., {Benni}
  P., {Foote} J., 2016, \mnras, 458, 3904

\bibitem[{{Rau} {et~al}\mbox{.}(2009){Rau}, {Kulkarni}, {Law}, {Bloom},
  {Ciardi}, {Djorgovski}, {Fox}, {Gal-Yam}, {Grillmair}, {Kasliwal}, {Nugent},
  {Ofek}, {Quimby}, {Reach}, {Shara}, {Bildsten}, {Cenko}, {Drake},
  {Filippenko}, {Helfand}, {Helou}, {Howell}, {Poznanski}, \&
  {Sullivan}}]{rau09}
{Rau} A. {et~al.}, 2009, \pasp, 121, 1334

\bibitem[{{Robin} {et~al}\mbox{.}(2003){Robin}, {Reyl{\'e}}, {Derri{\`e}re}, \&
  {Picaud}}]{Robi03}
{Robin} A.~C., {Reyl{\'e}} C., {Derri{\`e}re} S., {Picaud} S., 2003, \aap, 409,
  523

\bibitem[{{Roeser}, {Demleitner} \& {Schilbach}(2010){Roeser}, {Demleitner}, \&
  {Schilbach}}]{Roes10}
{Roeser} S., {Demleitner} M., {Schilbach} E., 2010, \aj, 139, 2440

\bibitem[{{Tonry} {et~al}\mbox{.}(2012){Tonry}, {Stubbs}, {Kilic},
  {Flewelling}, {Deacon}, {Chornock}, {Berger}, {Burgett}, {Chambers},
  {Kaiser}, {Kudritzki}, {Hodapp}, {Magnier}, {Morgan}, {Price}, \&
  {Wainscoat}}]{tonry12}
{Tonry} J.~L. {et~al.}, 2012, \apj, 745, 42

\bibitem[{{Tremblay} \& {Bergeron}(2009)}]{tremblay09}
{Tremblay} P.-E., {Bergeron} P., 2009, \apj, 696, 1755

\bibitem[{{Tremblay} {et~al}\mbox{.}(2011){Tremblay}, {Ludwig}, {Steffen},
  {Bergeron}, \& {Freytag}}]{Trem11}
{Tremblay} P.-E., {Ludwig} H.-G., {Steffen} M., {Bergeron} P., {Freytag} B.,
  2011, \aap, 531, L19

\bibitem[{{Tremblay} {et~al}\mbox{.}(2013){Tremblay}, {Ludwig}, {Steffen}, \&
  {Freytag}}]{tremblay13}
{Tremblay} P.-E., {Ludwig} H.-G., {Steffen} M., {Freytag} B., 2013, \aap, 559,
  A104

\bibitem[{{Udalski}(2003)}]{udalski03}
{Udalski} A., 2003, \actaa, 53, 291

\bibitem[{{Vanderburg} {et~al}\mbox{.}(2015){Vanderburg}, {Johnson},
  {Rappaport}, {Bieryla}, {Irwin}, {Lewis}, {Kipping}, {Brown}, {Dufour},
  {Ciardi}, {Angus}, {Schaefer}, {Latham}, {Charbonneau}, {Beichman},
  {Eastman}, {McCrady}, {Wittenmyer}, \& {Wright}}]{vanderburg15}
{Vanderburg} A. {et~al.}, 2015, \nat, 526, 546

\bibitem[{{Veras} {et~al}\mbox{.}(2013){Veras}, {Mustill}, {Bonsor}, \&
  {Wyatt}}]{Vera13}
{Veras} D., {Mustill} A.~J., {Bonsor} A., {Wyatt} M.~C., 2013, \mnras, 431,
  1686

\end{thebibliography}

\bsp    
\end{document}